\documentclass[pre,aps,reprint,superscriptaddress,titlepage,floatfix,showkeys]{revtex4-2} 
\usepackage{graphicx}
\usepackage{gensymb}
\usepackage{nicefrac}
\usepackage{amsfonts}
\usepackage[version=3]{mhchem}
\usepackage{amssymb}
\usepackage{amsmath}
\usepackage{subfigure}
\usepackage{multirow}
\usepackage{tabularx}
\usepackage{array}
\usepackage{units}
\usepackage{braket}
\usepackage{bm,times}
\usepackage{booktabs}
\usepackage{enumitem}
\usepackage{mathtools}
\usepackage{epstopdf}
\usepackage[dvipsnames, table]{xcolor}
\usepackage{threeparttable}
\usepackage[colorlinks = true, linkcolor = blue, urlcolor  = blue, citecolor = blue, anchorcolor = blue]{hyperref}

\def\etal{\it et al.\,}
\def\asio2{{\it a}-SiO$_2$} 
\def\vsio2{{\it v}-SiO$_2$} 
 
\def\asi{{\it a}-Si} 
\def\u2{$\langle u^2 \rangle$}

\def\AAI{{\AA}$^{-1}$}

\def\GSS{$G_{\text{\scriptsize SiSi}}(r)$}
\def\GSO{$G_{\text{\scriptsize SiO}}(r)$}
\def\GOO{$G_{\text{\scriptsize OO}}(r)$}

\def\FSS{$F_{\text{\scriptsize SiSi}}(Q)$}
\def\FSO{$F_{\text{\scriptsize SiO}}(Q)$}
\def\FOO{$F_{\text{\scriptsize OO}}(Q)$}

\def\bea{\begin{eqnarray}}
\def\eea{\end{eqnarray}}
\def\be{\begin{equation}}
\def\ee{\end{equation}}

\begin{document}

\title{
New insights into the origin of the first sharp 
diffraction peak in amorphous silica 
from an analysis of chemical and radial ordering
} 

\author{Parthapratim Biswas}
\email[Corresponding author:\,]{partha.biswas@usm.edu}
\affiliation{
Department of Physics and Astronomy, University of 
Southern Mississippi, Hattiesburg, MS 39406, USA
}

\author{Devilal Dahal}
\email{devilal.dahal@usm.edu}
\affiliation{
Department of Physics and Astronomy, University of Southern 
Mississippi, Hattiesburg, MS 39406, USA
}

\author{Stephen R. Elliott}
\affiliation{Physical and Theoretical Chemistry Laboratory,
University of Oxford, Oxford OX1 3QZ, United Kingdom}
%\email{stephen.elliott@chem.ox.ac.uk}

\affiliation{Department of Chemistry, University of Cambridge,
Cambridge CB2 1EW, United Kingdom}
\email{sre1@cam.ac.uk}

\begin{abstract}
The structural origin of the first sharp diffraction peak (FSDP) 
in amorphous silica is studied by analyzing chemical 
and radial ordering of silicon (Si) and oxygen (O) atoms 
in binary amorphous networks.  The study shows that 
the chemical order involving Si--O and O--O pairs play 
a major role in the formation of the FSDP in amorphous silica.  
This is supplemented by small contributions 
arising from the relatively weak Si--Si correlations in 
the Fourier space. A shell-by-shell analysis of the 
radial correlations between Si--Si, Si--O and O--O atoms 
in the network reveals that the position and the intensity 
of the FSDP are largely determined by atomic pair 
correlations originating from the first two/three radial 
shells on a length scale of about 5--8 {\AA}, whereas 
the fine structure of the intensity curve in the vicinity 
of the FSDP is perturbatively modified by atomic correlations 
arising from the radial shells beyond 8 {\AA}. The study 
leads to a simple mathematical relationship between the 
position of the radial peaks ($r_k$) in the partial pair-correlation 
functions and the diffraction peaks ($Q_k$) that can be 
used to obtain approximate positions of the FSDP and the 
principal peak. The results are complemented by numerical 
calculations and an accurate semi-analytical expression 
for the diffraction intensity obtained from the partial 
pair-correlation functions of amorphous silica for a given radial shell.  
\end{abstract}

%\keywords{Amorphous silica, First sharp diffraction peak, Reverse 
%Monte Carlo simulations, Chemical and radial ordering}

\maketitle
\section{Introduction} 
Amorphous silica ({\asio2}) is one of the most extensively 
studied noncrystalline solids. While the local structure 
of amorphous silica is well described in terms of the approximate 
tetrahedral arrangement of Si and O atoms leading to short-range 
ordering (SRO) in the network, the medium-range order (MRO) in 
{\asio2} is relatively less understood as far as the real-space 
ordering of the atoms is concerned~\cite{Wright:1991,Wright:1994,
SRE_NAT:1991,Salmon:2007,Kob:2020,Zhou:2023}. The SRO in 
network glasses is readily reflected in the pair-correlation 
functions (PCF) and the bond-angle distributions.  The appearance 
of the MRO in real space is much more subtle and difficult 
to gauge, however. 
In {\asio2}, the MRO is often associated with the 
relative position and orientation of Si[O$_4$]$_\frac{1}{2}$
tetrahedra in forming a continuous 
random network. In the presence of short-range ordering, 
driven by local chemistry and geometry, these structural 
motifs connect with the neighboring motifs in such 
a way that the resulting network exhibits atomic 
correlations on a nanometer length scale~\cite{Tanaka:2019}. 
Although these intermediate correlations 
reside in the PCF in a rather obscure way~\cite{Uhlherr:1995,Dahal:2021}, 
their presence is particularly evident in the Fourier 
space. The radial correlations in the PCF, 
extending up to a few nanometers in some glasses, 
can manifest in the form of a sharp diffraction 
peak in the region of 1--2 {\AAI} in the Fourier 
space. For {\asio2}, this leads to the presence of a sharp 
peak at $Q_1$ =  1.5 {\AAI}, also known as the first 
sharp diffraction peak (FSDP), which serves
as a key indicator of the presence of MRO in 
the network. 

The FSDP in glasses has been studied extensively via 
X-ray and neutron diffraction experiments~\cite{Inamura:2004,
Susman:1991t,Susman:1991p,Fischer:1988,Mei:2008,Crupi:2015,Zanatta:2014,Tanaka:1998,Shi:2019,Salmon:2023}, and 
computational modeling of glasses in simulation 
studies~\cite{Vashishta:1990,SRE_JPCM:1992,SRE_PRL:1991,
Sarnthein:1995,Uchino:2005,Dahal:2021,Koba:2023}. 
Although these studies have produced a wealth of structural 
information on glasses and established that the FSDP is a nearly universal 
feature of network glasses, a definitive understanding of 
the origin of the FSDP, and its thermodynamic behavior 
with respect to pressure and 
temperature~\cite{Inamura:2004,Susman:1991p,
Susman:1991t,Tan:1999,Inamura:2004,Zanatta:2014,Salmon:2023}, 
in terms of the real-space ordering of atoms on 
the medium-range length scale is yet to be achieved. 

In the past decades, several explanations for 
the possible origin of the FSDP in network glasses 
were proposed in the 
literature~\cite{Gaskell:1996,SRE_JPCM:1992,SRE_PRL:1991,
Price:1989a,Mei:2008,Massobrio:2001}.
Among these explanations, the viewpoints presented 
by Gaskell and Wallis~\cite{Gaskell:1996} and 
Elliott~\cite{SRE_PRL:1991,SRE_JPCM:1992} are 
most noteworthy. Following Gaskell and Wallis, 
the FSDP in network glasses is often attributed to the 
presence of diffuse quasi-Bragg planes, which are 
assumed to be separated by a distance of the order of $d_1=2\pi/Q_1$ 
in real space.  These quasi-Bragg planes are believed 
to originate from fluctuations of order, which produce 
a strong scattering in the region of 1--2~{\AAI} 
in most glasses. For {\asio2}, this translates into 
a value of $d_1 \approx$ 4 {\AA} and experimental data 
obtained from high-resolution electron microscope 
(HREM) images of {\asio2} appear to support the existence of these 
quasi-Bragg planes~\cite{Gaskell:1996}. 
The presence of quasi-Bragg planes in {\asio2} was 
also reported in a simulation study by Uchino {\etal}~\cite{Uchino:2005}, 
who employed a real-reciprocal space analysis of 
the pair-correlation function of {\asio2} by using a 
continuous wavelet transform technique.

An alternative view is provided by Elliott~\cite{SRE_NAT:1991,SRE_JPCM:1992}. In 
this approach, the chemical ordering of interstitial 
voids --  associated with cation-centered structural 
motifs or clusters -- in network glasses plays a central 
role in the formation of the FSDP. The author has 
shown that the FSDP in network glasses can be regarded as a prepeak in 
the concentration-concentration partial structure 
factor (obtained in the Bhatia-Thornton formalism~\cite{Bhatia:1970}), 
$S_{cc}(Q)$, originating from low-atomic-occupancy zones or 
voids in cluster+void networks.  A direct application of 
this cluster+void model leads to the position of 
the FSDP in {\asio2} near 1.52 {\AAI}~\cite{SRE_JPCM:1992}, 
which includes a corrective factor due to Bl{\'e}try~\cite{Bletry:1990}. 
The pressure and temperature dependence of the FSDP can be 
successfully described using this model for most 
AX$_2$-type (A = Si, Ge; X = O, S, Se) network glasses. 
Although some authors~\cite{Mei:2008} have disputed 
the applicability of the cluster+void model for {\asio2}, 
experimental studies on densified {\asio2} do indicate 
that interstitial voids play an important part in the 
formation of the FSDP and its behavior upon change of 
pressure and temperature~\cite{Zanatta:2014,Tan:1999}. 
It may be noted that these two seemingly different 
explanations are not totally unrelated to each other 
due to the presence of interstitial voids and hence 
the resulting fluctuations of the atomic density that 
can lead to strong scattering in the region 
of 1--2 {\AAI}. 

In this work, we employ a simple but deftly implemented 
approach in real space to understand and gain new insights 
on the origin and structure of the FSDP in {\asio2}.  
Although our work is focused on {\asio2}, the approach employed here 
is very general in nature and it can be applied to 
any network-forming amorphous solids, including amorphous 
silicon~\cite{Dahal:2021}. In our approach, we shall address the 
problem from a real-space point of view and demonstrate 
that the position of the FSDP (and the principal 
peak~\cite{SRE:1995} as well) can be explained from 
a knowledge of the partial PCFs and their relationship with 
the corresponding structure factors. In particular, we 
shall show that the FSDP corresponds to the minimal value 
of $Q$ for which radial correlations originating 
from distant atomic shells in real space interfere 
constructively to produce a strong intensity peak in 
the region of 1--2 {\AAI}. A direct 
consequence of this approach is that it can explain 
why certain atomic pairs (for example, Si--Si pairs
in {\asio2}) contribute very little to the FSDP, due to 
cancellation effects of contributions from neighboring 
radial shells. The key purpose of this study is to 
develop a quantitative approach for characterizing 
the radial contributions (of atomic correlations) 
from distant atomic shells to the FSDP, and to examine 
the relevant length scale(s) associated with these 
contributions.

The rest of the paper is presented as follows.  In Sec.~II, 
we discuss the computational method employed to generate the 
structure of {\asio2}, using a combination of Reverse Monte 
Carlo and first-principles simulations. 
Section~III discusses the results. 
Starting with the validation of the models, the positions of 
the FSDP and the principal peak are obtained by developing a 
simple ansatz from a knowledge of the partial PCFs. This is 
followed by a discussion on the origin of the FSDP and the 
principal peak, with particular emphasis on the contribution 
arising from individual radial shells of Si--Si, Si--O and 
O--O pairs in real space. The results are verified numerically 
and semi-analytically. The latter is achieved by deriving an 
accurate semi-analytical expression for the diffraction 
intensity originating from a given radial shell in the Gaussian 
approximation. The conclusions are presented in Sec.~IV.

\section{Computational method} 

\begin{figure}[t!]
\includegraphics[width=0.6\columnwidth]{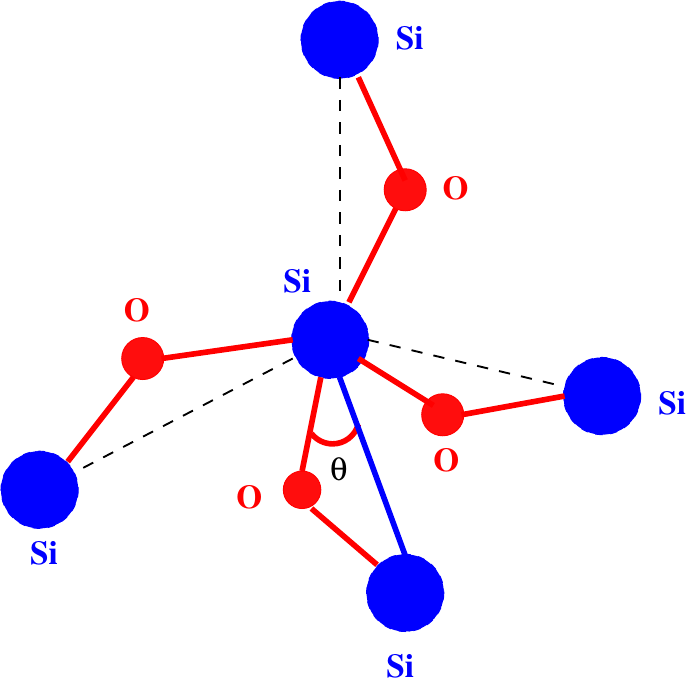}
\caption{
Construction of an initial configuration of {\asio2} from 
a tetrahedral model of amorphous silicon (blue) by introducing 
oxygen atoms (red) near the center of Si--Si bonds in 
the network. Each oxygen atom makes an angle $\theta$ of 
about 10--25{\degree} with its nearest Si--Si bond as 
indicated in the diagram.
}
\label{F1}
\end{figure}

The starting point of this study is to construct a model 
of {\asio2} using a combination of reverse Monte Carlo 
(RMC)~\cite{McGreevy:2001,Pandey:2015} and {\it ab initio} molecular 
dynamics (AIMD) simulations~\cite{Hutter:2009,Drabold:1990,Tafen:2003}. 
To this end, we begin with a 216-atom model of amorphous silicon 
({\asi}) with no coordination defects. This initial `seed' model is used as a framework 
structure, which can be augmented by adding oxygen atoms 
in the network. The generation of {\asio2} models in our 
approach thus consists of the following steps~\cite{Pandey:2015,Tafen:2003}: 

(i) The first step involves incorporation of oxygen atoms 
in a 216-atom {\asi} network. For a given Si--Si bond, 
an oxygen atom was placed close to (but not at) the center 
of the bond so that the resulting Si--O--Si bond angle 
lay between 130 to 160{\degree}. This was achieved by 
generating a random unit vector that made an angle 
$\theta$=10--25{\degree} with the Si--Si bond direction. 
An oxygen atom was then placed at a distance 
of approximately half of the Si--Si bond length along 
the unit vector. 
Care was taken to ensure that oxygen atoms were always 
placed either to the left or to the right of the four 
Si--Si bonds associated with a central silicon atom 
in order to construct an initial 
configuration as close as possible to the {\asio2} 
geometry. 
This is illustrated in Fig.~\ref{F1}, where four oxygen 
atoms can be seen to appear on the right of the Si--Si 
bonds (viewing clockwise from above) originating from 
the central Si atom to produce an approximate tetrahedral 
unit of Si[O$_4$]$_\frac{1}{2}$. The procedure was repeated 
for all Si--Si bonds in the network. 
The atomic positions in the resulting structure were 
then scaled by adjusting the length of the cubic 
simulation cell in order to match the mass density 
of the model with the experimental density of 
2.2~g.cm$^{-3}$~\cite{Mazurin:1983} of {\asio2}; 

(ii) Having included the correct stoichiometry, 
approximate chemical ordering, topology, and 
geometry of the {\asio2} structure in the network, 
the resulting model was subjected to RMC simulations 
in the second step. The RMC simulation proceeded 
by matching the experimental neutron-weighted total 
structure factor of {\asio2} with that calculated 
for the model. The simulation was conducted by incorporating 
an additional constraint on O--Si--O bond angles 
and their standard deviation so that the bond-angle 
distribution, imposed via selective addition 
of oxygen atoms in the first step, could not 
deviate considerably from the approximate tetrahedral 
distribution of O--Si--O bond angles. 
%The distribution 
%of the latter was ensured to have an average value 
%of 109.4{\degree} and a standard deviation of about 
%6--10{\degree}.  
The purpose of the RMC step here 
is to refine the overall radial structure of the model 
so that the (radial) distances between Si--O, O--O and 
Si--Si pairs of atoms are more or less consistent with those 
observed in diffraction experiments. 
Detailed discussions on the generation of amorphous structures using RMC 
simulations are given by McGreevy~\cite{McGreevy:2001} 
and others~\cite{Limbu:2020, Biswas:2004, Keen:1990, Opletal:2013}; 

(iii) The final step of the method involves low-temperature 
annealing or thermalization of the RMC-refined structure 
at 600~K, which is followed by total-energy relaxations using 
density functional theory (DFT). This is to ensure that the 
structural information enforced into the system in the 
last two steps is consistent with `thermodynamic' 
equilibrium or a strong minimum on the potential-energy 
surface of {\asio2}, as determined by total-energy 
and forces from DFT calculations. The annealing was done by conducting AIMD 
simulations in canonical ensembles at 600~K for a total 
simulation time of 5 ps using the code {\sc Siesta}~\cite{siesta}. 
{\sc Siesta} employs atom-centered numerical basis functions 
to solve the Kohn-Sham equation using the self-consistent 
field approximation within the framework of density functional 
theory (DFT). The norm-conserving pseudopotentials for silicon 
and oxygen atoms in the Troullier-Martins form~\cite{tm} were 
used to describe the electron-ion interactions, and the 
exchange-correlation energy of the system was computed in the 
generalized gradient approximation (GGA)~\cite{Perdew:1996}. 
The total-energy of the thermalized/annealed models was further 
minimized 
by employing the conjugate gradient (CG) method using {\it ab 
initio} forces and total-energy obtained from {\sc Siesta}. 
Throughout the calculations, we used double-zeta basis 
functions. The CG relaxation continued until the magnitude of 
the total force on each atom was less than or equal to 
0.01 eV/{\AA}.

To study the origin and structure of the FSDP and the principal 
peak in {\asio2} by analyzing the radial pair 
correlations between constituent atoms, we assume that the 
disordered system is isotropic and homogeneous in nature. 
%This is a reasonable assumption as long as the number of 
%defects are very low and the system does not exhibit significant 
%density fluctuations due to the presence of extended 
%inhomogeneities, such as voids in the network.   
Following Elliott~\cite{Elliott:1990}, the reduced scattering 
intensity ($I_r$) for such a system, consisting of $N$ atoms 
or scatterers, can be written as
\bea 
I_r & = & \frac{Q}{\langle f\rangle^2}\left[{\frac{I}{N}-\langle f^{2} 
\rangle}\right]  \notag \\ 
& = & \int_0^\infty 4\pi\rho_0 r [g(r)-1]\, \sin Qr \, dr, 
\label{E1} 
\eea
where $g(r)$ is the total atomic pair-correlation 
function (PCF), $\rho_{0}$ is the average number 
density of the system and $I$ is the total 
intensity. The symbol $\langle f^n\rangle$ 
(for $n$=1,2) stands for the concentration-weighted 
average value of the $n$-th moment of the scattering 
factor $f$.  Writing 
\be 
g(r) = \sum_{ij} \frac{c_i c_j f_i f_j}{\langle f \rangle^2} g_{ij}(r) 
= \sum_{ij} \omega^{\prime}_{ij}\, g_{ij}(r)    
\label{E2} 
\ee 
in terms of the partial PCFs, $g_{ij}(r)$, and noting 
that $c_i$ and $f_i$ are the atomic 
fraction and the scattering factor of atoms of type $i$, 
respectively, Eq.~(\ref{E1}) can be expressed as~\cite{Elliott:1990}
\bea 
\frac{I}{N \langle f^2 \rangle}- 1 & = & \sum_{ij}\frac{c_i c_j f_i f_j}{\langle f^2\rangle}\,[I_{ij}(Q)-1]  \notag \\
& = & \sum_{ij} \omega_{ij} [I_{ij}(Q) - 1]. 
\label{E3} 
\eea
In Eq.~(\ref{E3}), $I_{ij}$ is the  partial interference function 
\[ 
I_{ij}(Q) = 1 + \frac{1}{Q} \int_0^\infty 4\pi\rho_0\, r[g_{ij}(r)-1]\,\sin Qr\,dr
\] 
and $g_{ij}(r)$ is the partial PCF between atoms of type $i$ and type $j$. 
It may be noted that the coefficients 
$\omega_{ij}$ and $\omega^{\prime}_{ij}$ are close to each 
other but not identical because of the use of different 
denominators, ${\langle f^2 \rangle}$ and $\langle f \rangle^2$, 
respectively, in their definition, and 
$\sum_{ij} \omega^{\prime}_{ij} = 1$. 
For an elemental system, Eq.~(\ref{E3}) reduces to the well-known 
expression for the static structure factor
\be 
S(Q)= \frac{I}{Nf^2} = 1 + \frac{4\pi\rho_0}{Q} 
\int_0^\infty r[g(r)-1]\,\sin Qr\,dr. \notag 
\ee 
For our purpose, we rewrite Eq.~(\ref{E3}) as 
\be
S(Q)-1 = F(Q) = \sum_{ij} \omega_{ij} (I_{ij}(Q)-1) 
= \sum_{ij} F_{ij} \notag 
\ee 
where
\bea
F_{ij}(Q) &=& \frac{4\pi\rho_0\omega_{ij}}{Q} \int_0^{\infty} 
r\,[g_{ij}(r)-1] \sin Qr \, dr. 
\label{E4} 
\eea 
The upper limit of the integral in Eq.~(\ref{E4}) is generally 
truncated and replaced by a finite cutoff value of 
$R_c$. The latter is usually taken to be the half 
of the cubic simulation cell length or a value of $r$ for which 
$g_{ij}(r \ge R_c)$ = 0. The $R_c$ value for the present 
models is about 10.7 {\AA}. 
For neutron scattering, the expressions in this section are 
valid provided that one replaces the scattering factors, 
$f_i$, by the corresponding neutron-scattering lengths, $b_i$, 
for $s$-wave scattering~\cite{Cusack:1987,Bacon:1975}. The values 
of $b_{\text{\tiny Si}}$ and $b_{\text{\tiny O}}$ are given by 
4.149 and 5.803~fm, respectively.

\section{Results and Discussion} 
\subsection{Validation of structural models}

Before addressing the results, 
we briefly examine the static structure 
factor of the model and compare the results with those 
from experiments in order to validate the model used in 
our calculations. 
%Since the structure factor provides only two-body 
%atomic correlations, it is also necessary to inspect 
%the bond-angle distributions to obtain some information 
%regarding three-body correlations between atoms in 
%computer-generated models. 
For binary {\asio2}, it 
suffices to examine the distributions of Si--O--Si 
and O--Si--O bond angles, and the full structure factor 
for the purpose of validating a model. 

\begin{figure}[t!]
\includegraphics[width=0.7\columnwidth]{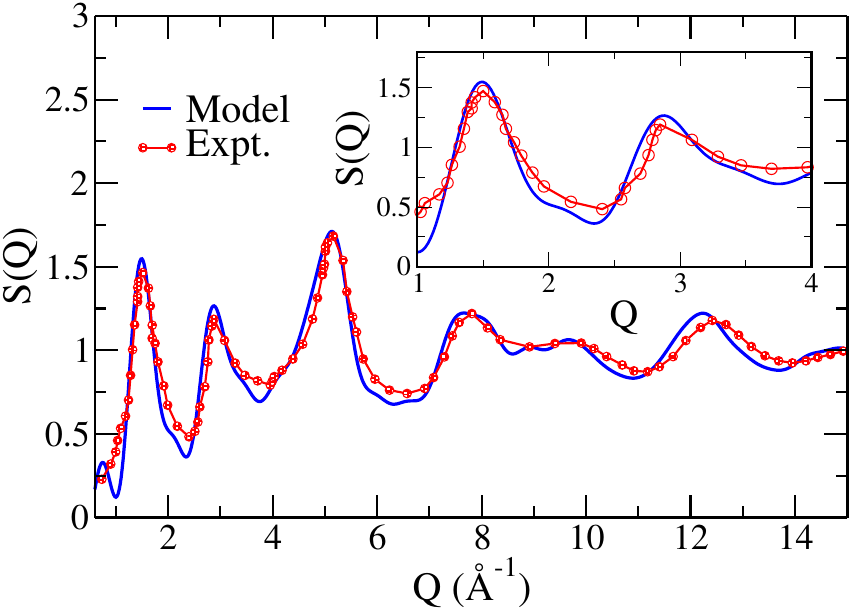}
\caption{
The neutron-weighted static structure factor of {\asio2}
obtained from averaging over two 648-atom models (blue line)
and experiments (red circles)~\cite{Susman:1991t}. The inset 
shows a magnified view of the FSDP and the principal peak, 
at 1.5 and 2.8 {\AAI}, respectively.
}
\label{F2}
\end{figure} 

Figure \ref{F2} shows the neutron-weighted static 
structure factor $S(Q)$ obtained from two 648-atom 
models of {\asio2}.  
Here, $S(Q)$ has been computed following the Faber-Ziman 
approach~\cite{Faber:1965} using Eq.~(\ref{E4}). The 
corresponding experimental data for {\asio2} from 
neutron diffraction measurements by Susman {\etal}~\cite{Susman:1991t} 
are also included in the figure. It is evident from the 
plots that the computed and experimental values agree 
well in the scattering-vector region from 0.6 to 15~{\AAI}. 
A small deviation below 1~{\AAI} can be attributed to 
finite-size effects on $G(r)$ obtained from small 
models. The presence of numerical noise in $G(r)$ 
at large distances can make $S(Q)$ particularly 
sensitive in the small-angle region of $Q <$ 1~{\AAI}~\cite{noise} 
for small finite-size models. 

The FSDP and the principal peak are found to be at 
1.5 and 2.8~{\AAI}, respectively, which are shown in 
the figure more closely as an inset along with 
their experimental counterparts. 
The position of the FSDP at 1.5 {\AAI} is also consistent with those 
from the X-ray diffraction measurements by Tan and 
Arndt~\cite{Tan:1999} and MD simulations reported in 
the literature by others~\cite{Vashishta:1990,Sarnthein:1995}. 
The average value of the Si--O bond length is found to be 
1.64$\pm$0.013 {\AA}, which is a bit larger than 
the experimental value of 1.61$\pm$0.09~{\AA}~\cite{Mei:2008,Johnson:1983} 
and 1.62$\pm$0.08{\AA} from first-principles MD simulations 
reported in the literature~\cite{Vashishta:1990,Tafen:2003,Sarnthein:1995}. 
Likewise, the average nearest-neighbor distances between 
Si--Si and O--O pairs are found to be 3.14$\pm$0.14~{\AA} 
and 2.63$\pm$0.06~{\AA}. 
These values compare well with the average Si--Si distance 
of 3.1~{\AA} and the average O--O distance of 2.64~{\AA} 
obtained from classical MD simulations~\cite{Vashishta:1990,ML:2022}, 
and 3.12~{\AA} and 2.65~{\AA}, respectively, from {\it ab initio} 
MD simulations~\cite{Tafen:2003}. The computed values 
are close to the experimental values 
of Si--Si and O--O distance of 3.08$\pm$0.1 and 
2.63$\pm$0.089 {\AA}, respectively, which were obtained 
by Johnson {\etal}~\cite{Johnson:1983} by assuming 
that the radial peaks are symmetric and can be fitted 
by a Gaussian function~\cite{Johnson:1983}.

\begin{figure}[t!]
\includegraphics[width=0.7\columnwidth]{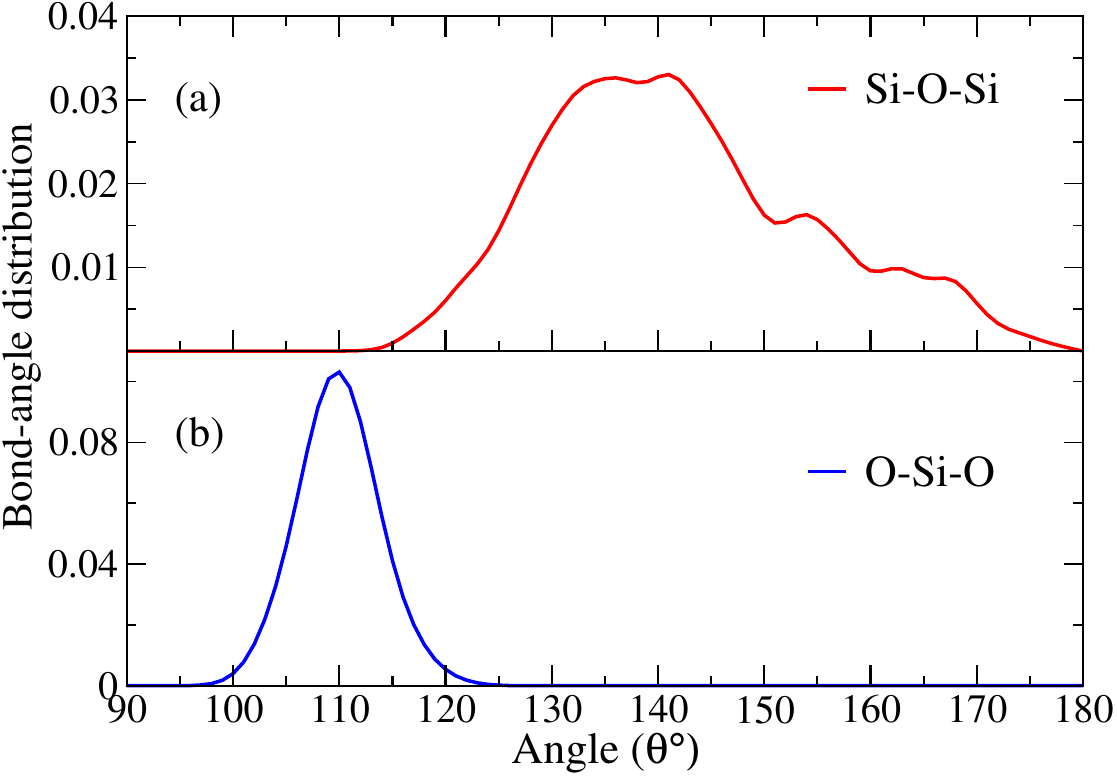}
\caption{
The bond-angle distributions associated with: (a) Si--O--Si 
angles and (b) O--Si--O angles. The average values of Si--O--Si 
and O--Si--O are 141.5{\degree} and 109.5{\degree}, 
respectively. The corresponding standard deviations are 
found to be 12.5{\degree} and 3.5{\degree}.
}
\label{F3}
\end{figure}

Turning to bond angles, the O--Si--O and Si--O--Si 
angles play a critical role in building the local 
structure and topology of {\asio2}, and the 
connectivity between SiO$_{\text{\scriptsize 4/2}}$ 
tetrahedra in the network.  
These two angles provide the relative shape and 
orientation of two neighboring SiO$_{\text{\scriptsize 4/2}}$ 
tetrahedra that form the structural basis of amorphous 
silica networks. The distributions of these angles are 
presented in Fig.~\ref{F3}.  The average value of 
O--Si--O angles in this work is found to be 109.5$\pm$3.5{\degree}.  
By contrast, the Si--O--Si angles exhibit a somewhat 
broader distribution, with an average value of 141.5$\pm$12.5{\degree}. 
These values match closely with those from earlier MD simulation 
studies~\cite{Sarnthein:1995, Tafen:2003, Vashishta:1990} 
and experiments~\cite{Mozzi:1969,Johnson:1983}. 
In particular, the Car-Parrinello simulations of {\asio2} 
using 72-atom models by Sarnthein et al.~\cite{Sarnthein:1995} 
obtained the average values for the O--Si--O and Si--O--Si 
angles to be 109$\pm$6{\degree} and 136$\pm$15{\degree}, 
respectively. The corresponding 
experimental values from X-ray and neutron diffraction
are reported to be about 109{\degree} 
and 140--150{\degree}~\cite{Mozzi:1969,Johnson:1983}. 

\subsection{Chemical order in {\asio2} and the FSDP} 

In studying the FSDP in network glasses, a question 
of considerable importance is to what
extent the presence of chemical ordering between 
constituent atoms/species can affect the position 
and the intensity of the diffraction peak. 
It is also necessary to ascertain the characteristic 
length scale of radial correlations that plays the 
most decisive role in forming the shape of the 
intensity curve near the FSDP. While it is an 
established fact that the FSDP originates from 
medium-range ordering on the length scale of 
5--10 {\AA}, a quantitative characterization 
of the contributions arising from different radial 
shells and atomic species is still missing in the 
literature. For a binary {\asio2} system, this 
entails examining the role of radial atomic correlations 
that originate from chemically ordered Si--Si, Si--O 
and O--O pairs in forming the FSDP.  To address 
this question, we write 
\[
F(Q) = \sum_{ij} \omega_{ij} (I_{ij}-1) = \sum_{ij} F_{ij}(Q)
\] 
and 
\be 
F_{ij}(Q) = \frac{1}{Q} \int_0^{R_c} G_{ij}(r) \sin Qr\, dr, 
\label{E5}
\ee
where $G_{ij}(r)=4\pi\rho_0\,\omega_{ij}\,r[g_{ij}(r)-1]$ 
is the neutron-weighted reduced atomic pair-correlation 
function. Equation (\ref{E5}) provides a convenient starting point 
to determine the contribution 
to the total FSDP arising from $F_{ij}(Q)$, or its Fourier counterpart 
$G_{ij}(r)$ in real space. 

\begin{figure*}[t!]
\includegraphics[width=0.33\linewidth]{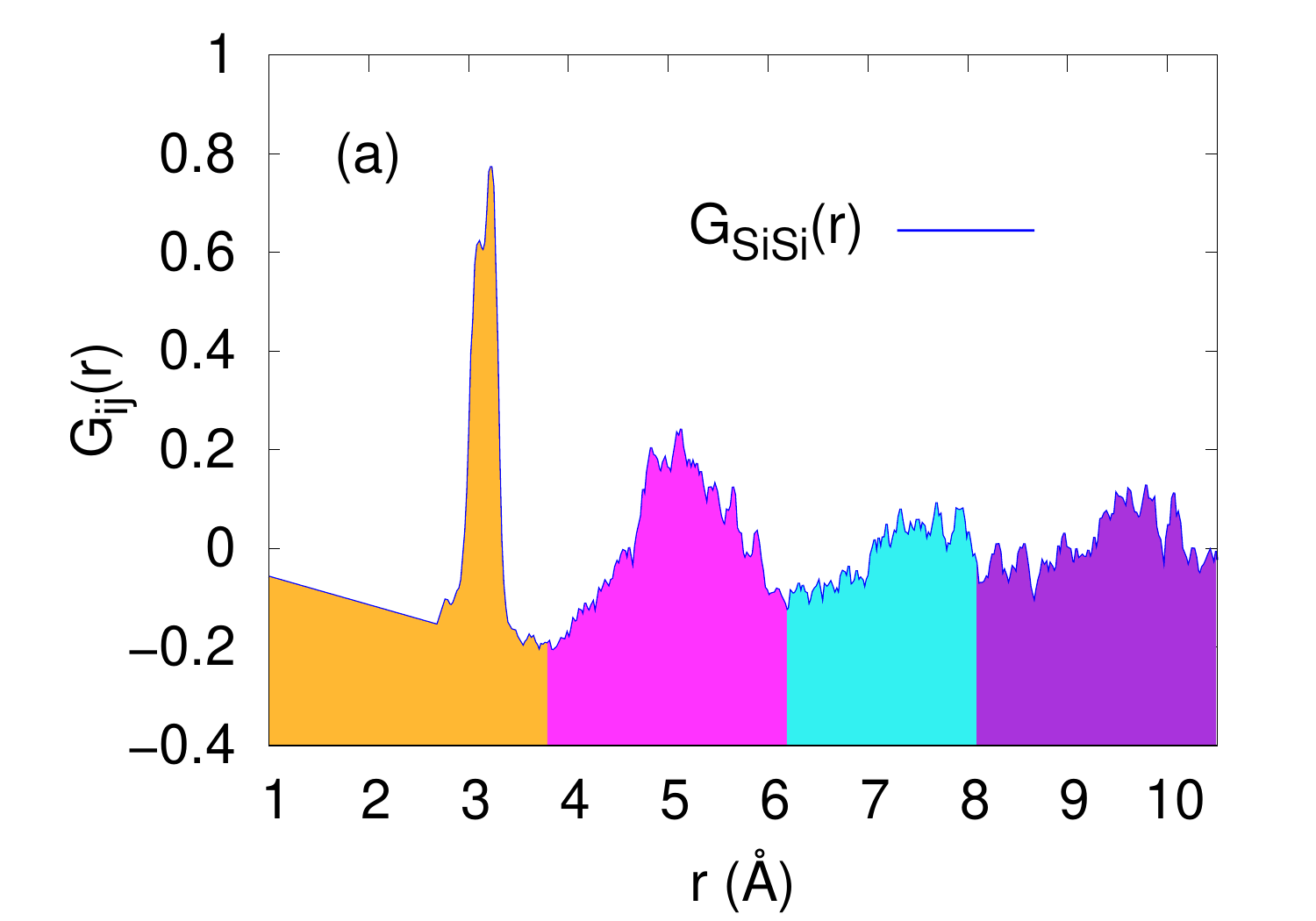}
\includegraphics[width=0.33\linewidth]{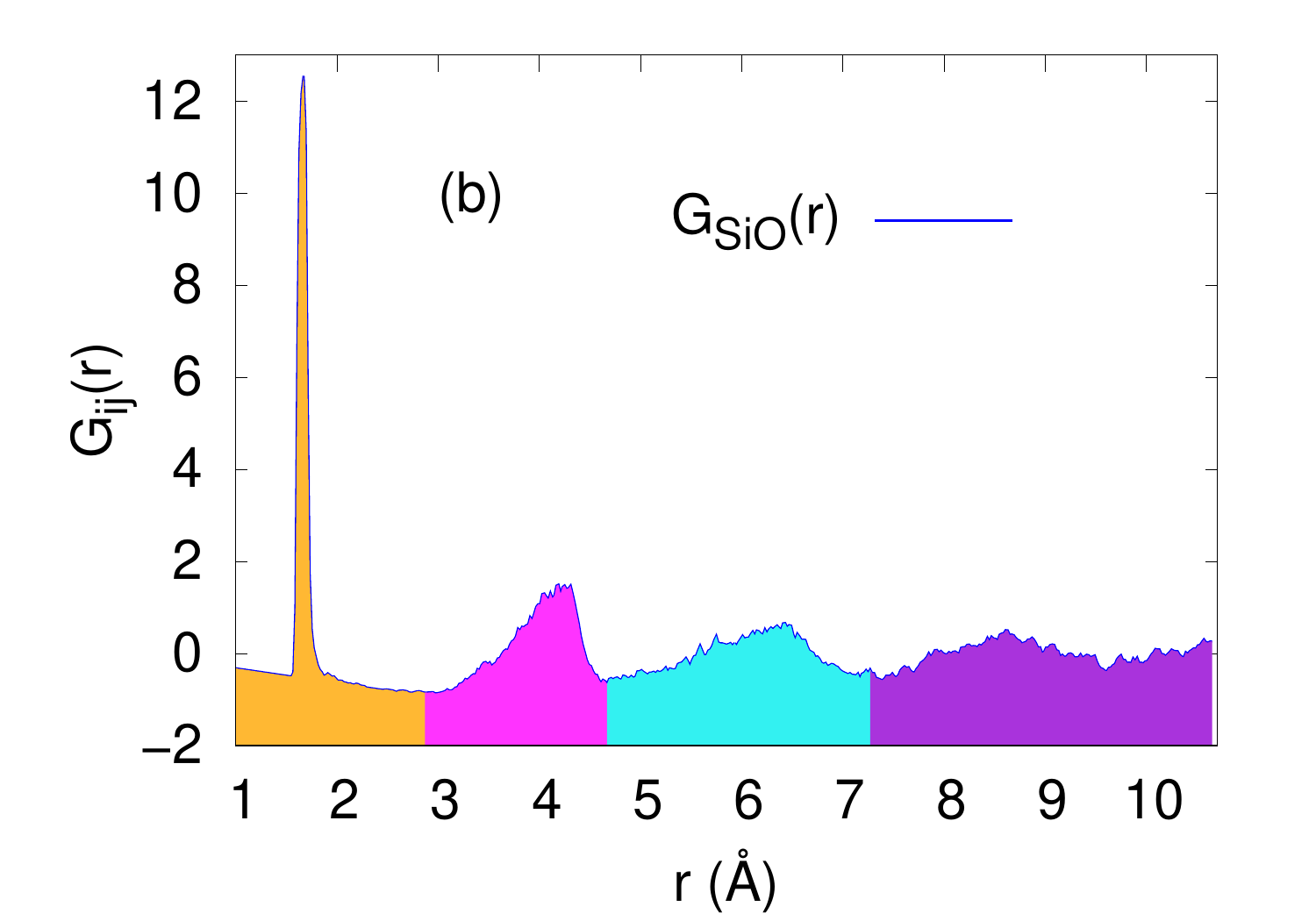}
\includegraphics[width=0.33\linewidth]{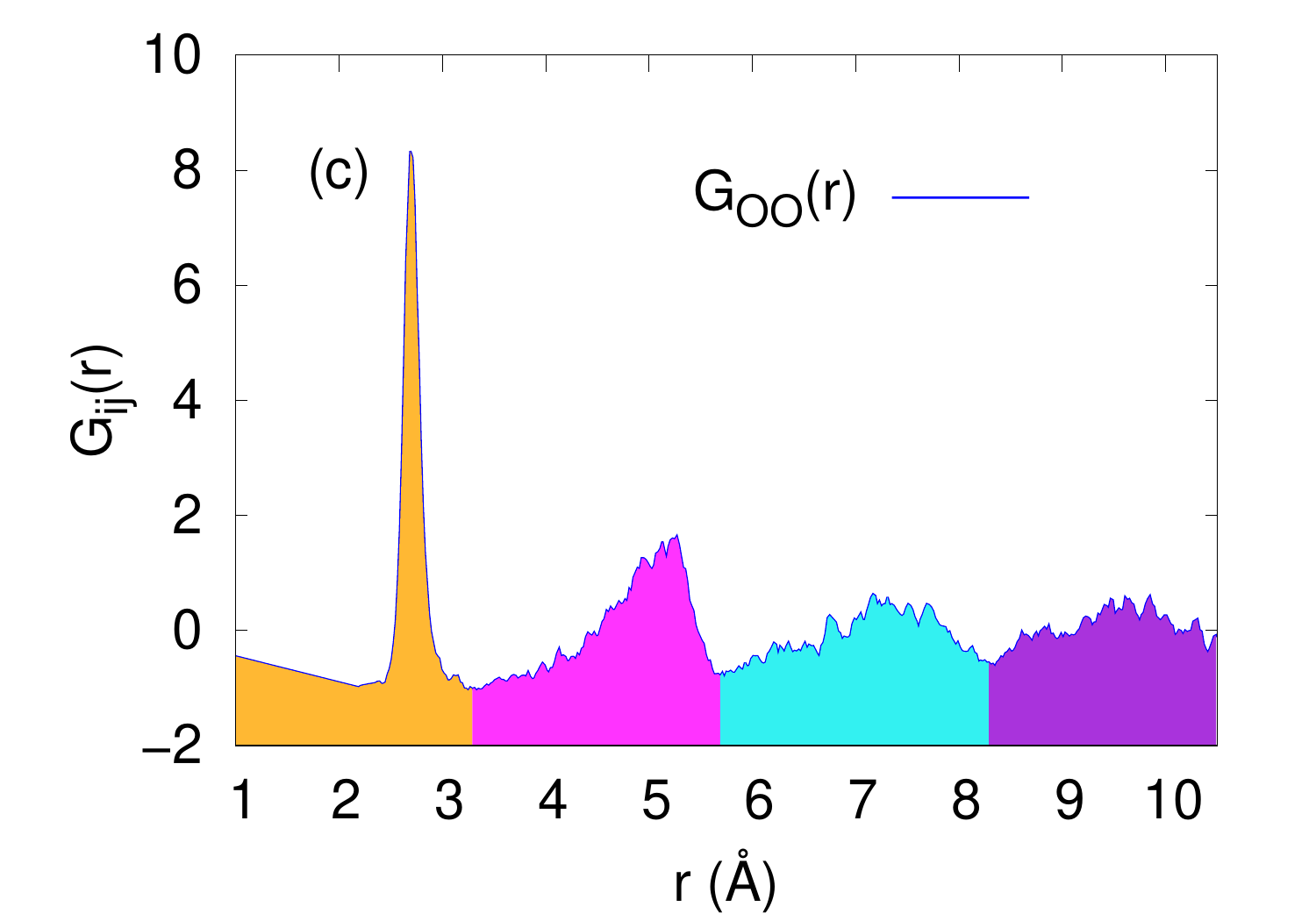}
\caption{
The neutron-weighted partial reduced PCFs, $G_{ij}(r)$, for: (a)
Si--Si, (b) Si--O and (c) O--O pairs. The first four shells and their
radial extent are shown in different colors. The magnitude of 
Si--Si pair correlations can be seen to be significantly smaller 
than their Si--O and O--O counterparts, by a factor of about 10.
}
\label{F4}
\end{figure*}

\begin{figure}[t!]
\includegraphics[width=0.7\columnwidth]{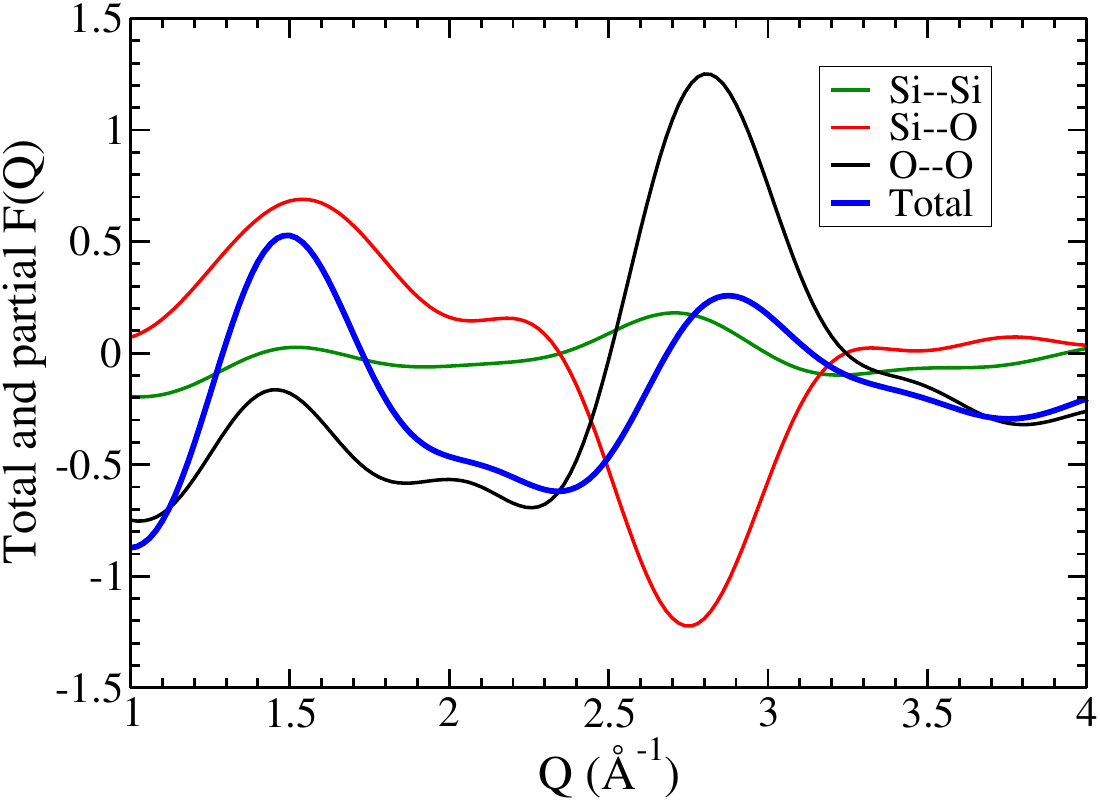}
\caption{
Neutron-weighted total $F(Q)$ (blue) and its 
three partial components, {\FSS}, {\FOO} and {\FSO}, 
originating from Si--Si (green), O--O (black) 
and Si--O (red) radial pair correlations, 
respectively. 
The FSDP and the principal peak of $F(Q)$ are
located at 1.5 and 2.8~{\AAI}, respectively.
}
\label{F5}
\end{figure}

Figure \ref{F4} shows the three neutron-weighted partial 
PCFs, $G_{ij}(r)$, for {\asio2}. It is apparent from 
the plots that neutron-weighted Si--Si correlations are considerably 
weaker than their Si--O and O--O counterparts, with the 
first two peaks of $G_{\text{\scriptsize SiSi}}(r)$ 
being about ten times smaller than the corresponding values of 
$G_{\text{\scriptsize SiO}}(r)$ and $G_{\text{\scriptsize OO}}(r)$. 
In view of this, it is not inapposite to surmise that the Si--Si 
correlations may not play a significant role in determining the 
overall structure of the FSDP. This surmise can 
be verified by computing $F_{ij}(Q)$ from Eq.~(\ref{E5}). 
Figure \ref{F5} shows the contribution to $F(Q)$ 
from its three partial components $F_{ij}(Q)$. 
The major contribution to the FSDP at 1.5 {\AAI} 
can be seen to originate from the Si--O correlations, which 
are followed by the O--O correlations in real space. This 
statement applies to the principal peak near 2.8 {\AAI} as 
well. The presence of weak Si--Si correlations, relative 
to the magnitude of its Si--O and O--O counterparts, 
produces an almost flat structure factor in the Fourier 
space, with two small bulges near 1.5 and 2.7 {\AAI}. 
An analysis of Eq.~(\ref{E5}), with the aid of Fig.~\ref{F4} 
in the next section, will reveal that the first bulge in 
$F_{\text{\scriptsize SiSi}}(Q)$ near the FSDP arises from 
the second radial shell of Si--Si correlations (depicted 
in pink-purple color in Fig.~\ref{F4}a), whereas the principal 
peak in $F_{\text{\scriptsize SiSi}}(Q)$ gets its 
contribution from both the first and second radial shells 
of silicon, extending as far as 6 {\AA}. 

By contrast, the position of the principal peak in $F(Q)$ 
(in Fig.~\ref{F5}) is determined by atomic correlations 
originating from {\GOO}, {\GSO} and {\GSS} as follows. 
The contribution from {\GOO} produces a strong peak 
or maximum near 2.8~{\AAI} in {\FOO}, whereas {\GSO} 
yields an equally strong minimum near 2.76~{\AAI} in 
{\FSO}. These two contributions 
effectively cancel out each other. The resulting intensity 
obtained from the sum of {\FOO}, {\FSO} and a 
small positive contribution from {\FSS} near 2.74~{\AAI} 
produces the principal peak near 2.8~{\AAI} in $F(Q)$. 
It thus appears that the chemical ordering of 
Si--O and O--O pairs plays an important role 
by competing with each other in determining the final 
position and intensity of the principal peak in 
amorphous silica. On the other hand, the 
Si--Si pairs provide a small but nontrivial contribution 
to those originating from Si--O and O--O pairs. 
Below, we discuss this in detail by developing an 
ansatz to obtain the approximate positions of the FSDP 
and the principal peak from Eq.~(\ref{E5}).  

\subsection{Origin of the FSDP and the principal peak in {\asio2}} 
The structural origin of the FSDP and the principal peak in {\asio2} at 
1.5 and 2.8 {\AAI}, respectively, can be intuitively understood 
from the approximate behavior of Eq.~(\ref{E5}) and a knowledge 
of the partial PCFs. To this end, we rewrite Eq.~(\ref{E5}) as 
\be
F_{ij}(Q) = \int_0^{R_c} rG_{ij}(r)\,\left[\frac{\sin Qr}{Qr}\right]\, dr \\
\label{E6}
\ee
and postulate -- following the saddle-point approximation 
of the integral in Eq.~(\ref{E6}) -- that the maxima (minima) 
of $F_{ij}(Q)$ should roughly correspond to those values 
of $Q$ for which $\sin Qr/Qr$ is maximum (minimum), where 
the values of $r$ are given by the dominant peaks in 
$rG_{ij}(r)$. Noting that the position of peaks in 
$G(r)$ and $rG(r)$ are very close to each other~\cite{peak}, 
and that the maxima (minima) of $(\sin x)/x$ can be expressed 
in terms of the maxima (minima) of $\sin x$, one can write in 
the first-order approximation
\be 
Q_k \approx \frac{(4n \pm 1)\pi}{2r_k} - \frac{2}{(4n \pm 1)\pi r_k}, \quad n = 1, 2, 3, \ldots 
\label{E7}
\ee 
where $r_k$ corresponds to the $k$-th peak in $G_{ij}(r)$ and 
the plus (minus) sign applies to the maxima (minima). 
This can be shown by solving the optimality condition,
$\cos x = (\sin x)/x$, ($x = Qr$) for the maxima/minima of
$(\sin x)/x$ in terms of the solutions of $\sin x = \pm 1$
or $x_n = (4n \pm 1) \pi/2$ for $n=0, 1, 2, \ldots$
Writing $x = x_n + x^{\prime}$, where $x^{\prime}$ is
a small correction, one arrives at the first-order
correction $x^{\prime} = -1/x_n$ via power series
expansion of $\sin x^\prime$ and $\cos x^\prime$ for
small $x^{\prime}$ in the expression for the optimality
condition.
For practical purposes, the second term in Eq.~(\ref{E7}), 
which is of the order of 0.1~{\AAI}~\cite{corr} 
for {\asio2}, can be ignored and the approximate values 
of $Q_k$ then read
\be 
Q_k \approx \frac{(4n \pm 1)\pi}{2r_k}, \quad n = 1, 2, 3, \ldots 
\label{E8}
\ee 

The validity of the result above rests on the assumption that the 
first few peaks in $G(r)$ are well defined and that they 
progressively decay so that the contribution from distant radial 
shells/peaks have a decreasing influence on the resultant FSDP 
and the principal peak in $F(Q)$. For elemental systems, 
such as {\asi}, it has been shown recently that Eq.~(\ref{E8}) 
provides a good estimate of the positions of the FSDP and the 
principal peak~\cite{Dahal:2021}. However, complications can arise 
for multinary systems where one must take into account 
not only the contributions from the distant radial shells 
but also those that originate from different partial components 
of $G(r)$. Since the contributions from the first 
few peaks of different $G_{ij}(r)$ could be similar in 
magnitude but out of phase to varying degrees, the resultant 
FSDP and the principal peak may or may not appear in the 
vicinity of $Q_k$ as determined by Eq.~(\ref{E8}), due to the 
cancellation of out-of-phase contributions. 
This is apparent in Fig.~\ref{F5}, where the 
position of the principal peak is determined by the sum of 
the intensity from the neutron-weighted Si--Si, Si--O and O--O 
pair correlations in the Fourier space. 
The intensity from Si--O and O--O correlations 
cancel each other out for the most part near the principal 
peak, leaving behind a resulting nontrivial 
part that combines with the small contribution from 
Si--Si correlations to determine the final position 
and the shape of the principal peak at 2.8 {\AAI}. 
This observation suggests 
that a knowledge of the intensity of $F_{ij}(Q)$ near the 
FSDP and the principal peak can be very useful in determining the 
exact final position of the peak in multinary systems. We 
shall delve into this point in the next section. For now, 
we focus on Eq.~(\ref{E8}) to obtain the approximate 
positions of the FSDP and the principal peak of {\asio2} 
with the aid of the neutron-weighted 
partial PCFs shown in Fig.~\ref{F4}.

Starting with the Si--Si PCF, one notes from Fig.~\ref{F4} 
that the positions of the first three peaks (at $r_k$) 
are at 3.14, 5.1 and 7.7~{\AA}, respectively.  Following 
Eq.~(\ref{E8}), the first radial peak at 3.14~{\AA} is 
expected to produce a peak at 2.5~{\AAI} (for $n$ = 1) 
in the Si--Si partial structure factor. Likewise, the 
second radial peak at 5.1~{\AA} should produce peaks 
at 1.54 and 2.77~{\AAI} for $n$ = 1,2, respectively. 
A similar calculation for the third radial peak at 
7.7~{\AA} suggests that a peak should appear at 
1.83 {\AAI} (for $n$=2) and 2.65~{\AAI} (for $n$=3). 
An examination of the Si--Si partial structure factor 
in Fig.~\ref{F5} does confirm the presence of peaks 
near 1.5 and 2.7 {\AAI}. In view of this observation, it 
is apposite to conclude that the Si--Si principal peak near 2.7~{\AAI} in 
Fig.~\ref{F5} gets its contribution mostly from the second 
and third Si--Si radial shells, whereas only the second 
shell contributes to the FSDP near 1.5~{\AAI}. The 
weak intensity associated with these peaks can be 
attributed to the relatively small values of 
neutron-weighted {\GSS} (compared 
to its Si--O and O--O counterparts), the largest 
absolute value of which over the entire radial range 
is found to be less than unity in Fig.~\ref{F4}a. 

The foregoing analysis applies to the O--O pairs 
as well.  Figure~\ref{F4}c shows that the first three peaks 
in {\GOO} appear at 2.65, 5.2 and 7.2~{\AA}.  Following 
the same reasoning as before, the 
first radial peak at 2.65~{\AA} should give rise to a peak at 2.96~{\AAI} 
(for $n$ = 1) and the second radial peak at 5.2~{\AA} 
leads to 1.51 and 2.72~{\AAI} in {\FOO} for $n$=1,2, 
respectively. Similarly, the third peak at 7.2~{\AA} 
is expected to produce peaks at 1.96 and 2.83~{\AAI} 
for $n$ = 2,3, respectively.  A comparison of the 
results with {\FOO} in Fig.~\ref{F5} shows that the 
suggested peaks appear near 1.5 and 2.8 {\AAI}, including 
a weak quasi-peak near 2~{\AAI}. Table \ref{tab1} lists 
the approximate positions of the FSDP and the principal 
peak of {\FSS} and {\FOO} obtained from using 
Eq.~(\ref{E8}) and the location of the first 
three radial peaks of {\GSS} and {\GOO}.  The values 
of $Q$ that contribute to the FSDP and the principal 
peak are highlighted in Table \ref{tab1} in orange 
and green colors, respectively. 

The case for Si--O correlations is a bit confusing, however.  
In contrast to {\FSS} and {\FOO}, there is no principal 
peak or maximum in {\FSO} but a minimum near 2.8~{\AAI} (see Fig.~\ref{F5}). 
The preceding argument works well to determine this minimum 
using the position of the {\em minima} obtained from 
Eq.~(\ref{E8}). In particular, the first three peaks of {\GSO}, 
appearing at 1.65, 4.27 and 6.4 {\AA}, are expected 
to produce three minima at 2.86~{\AAI} (for $n$ = 1), 2.58~{\AAI} 
(for $n$ = 2) and 2.7~{\AAI} (for $n$ = 3), respectively. 
The combined effect of these minima is expected to 
reflect in {\FSO} by producing a strong minimum near 
2.8 {\AAI}. Figure \ref{F5} does corroborate this 
expectation. We shall further confirm this observation 
using numerical results in Sec.~IIID. 
However, the argument appears to fail in locating the FSDP 
of {\FSO} at 1.5~{\AAI}. Equation (\ref{E8}) suggests that the 
second radial shell should produce a peak at 1.84~{\AAI} 
(for $n$ = 1) and for the third shell at 1.23~{\AAI} 
(for $n$=1). But none of these positions is 
close enough to the observed peak at 1.5~{\AAI} in 
Fig.~\ref{F5}. Should we consider the fourth radial shell 
(with a peak at 8.6~{\AA} in Fig.~\ref{F4}b), we 
get peaks in {\FSO} at 0.91~{\AAI} (for $n$=1) and 1.64~{\AAI} (for $n$=2). 
But without a knowledge of the intensity of these peaks, 
it is not clear whether the peaks would coalesce into 
a single peak at 1.5~{\AAI} or not. The approximate 
positions of the maxima and minima of {\FSO} obtained 
from using Eq.~(\ref{E8}) for the first four radial 
peaks are listed in Table~\ref{tab2}.

\begin{table}
\caption{
Approximate positions ($Q$) of the FSDP (orange cells) and 
the principal peak (green cells) obtained from the first 
three radial peaks ($r_i$) of {\GSS} and {\GOO} using 
Eq.~(\ref{E8}). For a given radial peak ($r_i$) and 
$n$, and $K = (4n+1)\pi/2$, the values of $Q = K/r_i$ 
are listed below in {\AAI}.  
\vspace*{0.1cm}
\label{tab1}
}
\scalebox{1.0}{
\begin{tabular}{|c|c|ccc|ccc|}
\hline
\multicolumn{1}{|c|}{$n$} & {$K$} &
\multicolumn{3}{c|}{Si--Si ($r_i$)} &
\multicolumn{3}{c|}{O--O ($r_i$)} \\
\cline{3-8}
 & & 3.14 \AA \: \:  & 5.1 \AA \: \: & 7.7 \AA \: \: & 2.65 \AA \: \: & 5.2 \AA  \: \:  & 7.2 \AA \: \: \\
\hline
\: 1 \: & $5\pi/2$ & 2.5 & \cellcolor{Peach} 1.54 & 1.02 & \cellcolor{YellowGreen} 2.96 & \cellcolor{Peach} 1.51 & 1.1\\[4pt]
\: 2 \: & $9\pi/2$ & 4.5 & \cellcolor{YellowGreen} 2.77 & 1.83 & 5.33 & \cellcolor{YellowGreen} 2.72 & 1.96\\[4pt]
\: 3 \: & $13\pi/2$& 6.5 & 4.0 & \cellcolor{YellowGreen} 2.65 & 7.7 & 3.93 & \cellcolor{YellowGreen} 2.83\\[4pt]
\hline
\end{tabular}}
\end{table}

The discrepancy observed above concerning the position of 
the estimated FSDP in {\FSO} cannot be satisfactorily 
explained using the qualitative argument [based on Eq.~(\ref{E8})] 
presented so far in this section. At this point, it 
suffices to mention that the anomaly can be resolved 
by noting that the peak at 1.23~{\AAI} originating 
from the third shell interferes 
constructively with those at 1.84~{\AAI} from the second shell 
and at 1.64~{\AAI} from the fourth shell (see Table 
\ref{tab2}) to produce the FSDP at 1.5~{\AAI}. 
In the next two sections, we address this issue by 
numerical and semi-analytical calculations, and 
examine the observed behavior of $F_{ij}(Q)$ with 
reference to neutron-weighted atomic pair 
correlations arising from individual radial shells 
of the partial PCFs of {\asio2}.

\begin{table}
\caption{
The approximate location of the maxima ($Q_{+}$) and 
minima ($Q_{-}$) in {\FSO} obtained from the first 
four shells using Eq.~(\ref{E8}). $K_{(+,-)}$ are 
given by $(4n\pm1)\pi/2$ and 
$Q_{\pm} = K_{(+,-)}/r_i$. The maxima and minima 
contributing to the FSDP and the principal minima in 
{\FSO} are indicated in orange and green cells, 
respectively.  
\vspace*{0.05cm}
\label{tab2}
}
\scalebox{0.87}{
\begin{tabular}{|c|c|cccc|cccc|}
\hline
\multicolumn{1}{|c|}{} & &
\multicolumn{8}{c|} {Si -- O ($r_i$)} \\
\cline{3-10}
\: $n$ \: & $K_{(+, -)}$ & 1.65 \AA \:   & 4.27 \AA \:   & 6.4 \AA \: & 8.6 \AA   & 1.65 \AA \:  & 4.27 \AA \:  & 6.4 \AA \: & 8.6 \AA \: \\
\cline{3-10}
& & \multicolumn{4}{c|}{$Q_+$ (maxima) } & \multicolumn{4}{c|}{$Q_{-}$ (minima)} \\
\hline
\: 1 \: & $\frac{(5,3)\pi}{2}$  & 4.76  & \cellcolor{Peach} 1.84 & \cellcolor{Peach} 1.23 & 0.91 & \cellcolor{YellowGreen} 2.86  & 1.1  & 0.74 & 0.54 \\[4pt]
\: 2 \: & $\frac{(9, 7)\pi}{2}$ & 8.57  & 3.31 & 2.21 & \cellcolor{Peach} 1.64 &  6.66  & \cellcolor{YellowGreen} 2.58  & 1.72 & 1.28 \\[4pt]
\: 3 \:  & $\frac{(13,11)\pi}{2}$& 12.37 & 4.78 & 3.19 & 2.37 & 10.47 & 4   & \cellcolor{YellowGreen} 2.7 & 2 \\[4pt]
\: 4 \:  & $\frac{(17,15)\pi}{2}$& 16.18 & 6.25 & 4.17 & 3.1 &  14.28 & 5.52   & 3.68 & \cellcolor{YellowGreen} 2.74 \\[4pt]
\hline
\end{tabular}}
\end{table}

\subsection{Relation between radial correlations, the 
FSDP and principal peak} 

In the preceding section, we have seen that the position
of the FSDP and the principal peak can be obtained -- except 
for the FSDP in {\FSO} -- from a heuristic argument based 
on the qualitative behavior of the integral in Eq.~(\ref{E6}) 
and the position of the radial peaks in $G_{ij}(r)$. 
The argument tacitly assumed that the 
function $G_{ij}(r)$ can be locally replaced by a sufficiently 
narrow pair-distribution function or a $\delta$ function at 
the maximum of each radial shell in the zeroth-order approximation. This 
consideration leads to a reasonable estimate of the position of 
the maxima/minima in the corresponding partial structure factor, 
but not the intensity of the peaks/dips. We now examine 
the accuracy of the position of maxima/minima obtained 
earlier by numerical calculations and study the radial 
contribution originating from the individual shells of 
$G_{ij}(r)$.

To obtain the shell-by-shell contribution from the real-space 
PCFs, it is convenient to write 
\bea 
F_{ij}(Q) &=& \sum_{\alpha=1}^s F^{\alpha}_{ij}(Q; R^{ij}_{\alpha}, 
R^{ij}_{\alpha+1}) \notag \\
&=& \frac{1}{Q} \, \sum_{\alpha=1}^s \int_{R^{ij}_\alpha}^{R^{ij}_{\alpha + 1}} 
\,G_{ij}(r)\,\sin Qr\,dr. 
\label{E9}
\eea
Here, the Greek index $\alpha$ indicates the shell number 
and the pair of lengths $(R^{ij}_\alpha, R^{ij}_{\alpha +1})$ 
gives the radial range of the $\alpha$-th shell for the 
$ij$-th partial PCF, $G_{ij}(r)$. The set $\{R^{ij}_\alpha\}$
is so chosen that it spans the entire radial range of the 
corresponding PCF from 0 to $R_c$. Table \ref{tab3} 
lists the pair of values ($R^{ij}_{\alpha}, R^{ij}_{\alpha + 1}$) 
for the $\alpha$-th shell of $G_{ij}(r)$. We now use these 
values of $\{R^{ij}_{\alpha}\}$ to calculate the radial 
contribution of the partial PCFs to the intensity of the 
corresponding $F_{ij}(Q)$ from individual radial shells. 
Our discussion is mostly confined to Si--O and O--O pair correlations 
as these two pairs have been found to play a key role in 
forming the FSDP and the principal peak (cf.~Fig.~\ref{F5}). 
This is followed by a brief mention of Si--Si correlations, 
which provide small corrections to the intensity of the 
FSDP/principal peak of the full $F(Q)$. 

\begin{table}
\caption{
Radial extents of the first four shells in the 
partial PCF of Si--Si, Si--O and O--O, as shown 
in Fig.~\ref{F4}. The pair of numbers within 
brackets below indicates the range of the $\alpha$-th 
radial shell in~{\AA}. 
}
\begin{ruledtabular}
\begin{tabular} {c|cc|cc|cc|cc}
\multicolumn{1} {c|} {PCF} &
\multicolumn{8} {c }{Shell number} \\
\cline{2-9}
\multicolumn{1} {c|}{$\downarrow$} &
\multicolumn{2} {c|}{$\alpha$ = 1}  &
\multicolumn{2} {c|}{$\alpha$ = 2}  &
\multicolumn{2} {c|}{$\alpha$ = 3}  &
\multicolumn{2} {c}{$\alpha$ = 4}  \\
\hline
Si--Si & (0, 3.8) & & (3.8, 6.2) & & (6.2, 8.1) &  & (8.1, 10.7) &  \\[3pt]
Si--O  & (0, 2.4)  & & (2.4, 4.6) & & (4.6, 7.4) &  & (7.4, 10.7) &  \\[3pt]
O--O   & (0, 3.3) & & (3.3, 5.7) & & (5.7, 8.3) &  & (8.3, 10.7) &  \\[3pt]
%\hline
\end{tabular}
\end{ruledtabular}
\label{tab3}
\end{table}

\begin{figure}[t!]
\includegraphics[width=0.7\columnwidth]{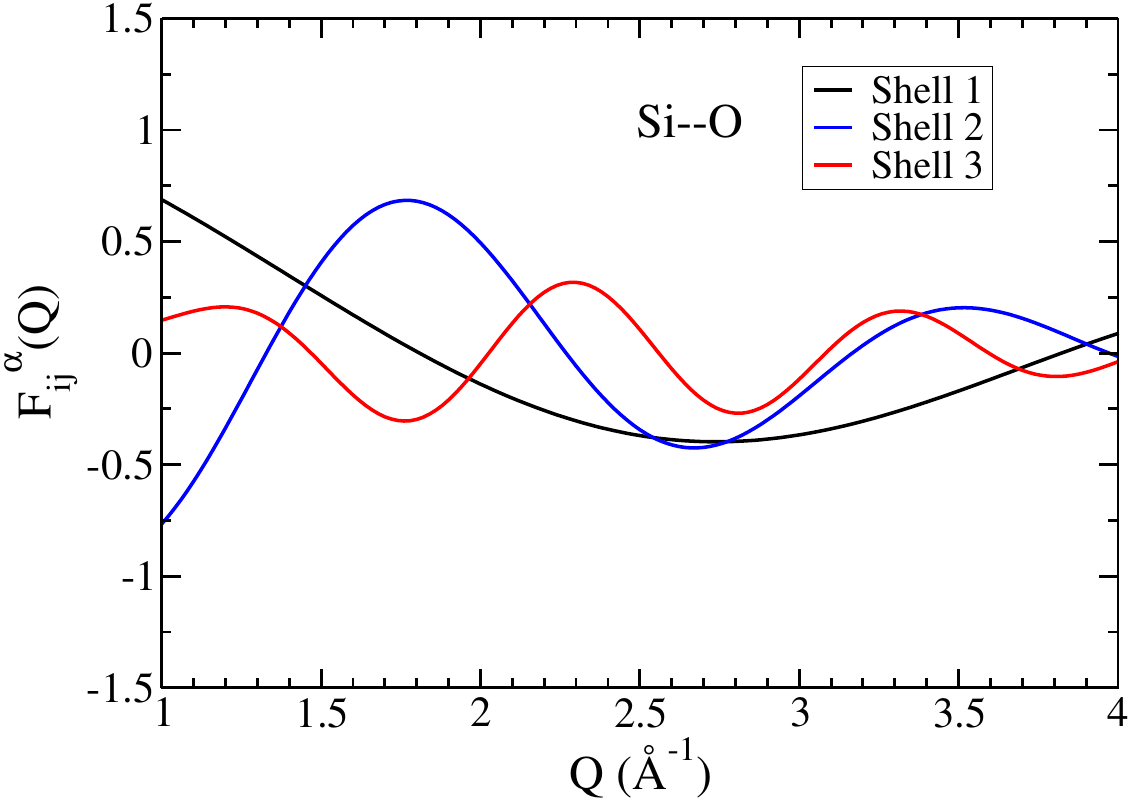}
\caption{
The partial structure factors, $F_{ij}^\alpha(Q)$, 
obtained from the first three radial shells (for 
$\alpha$ = 1, 2 and 3) of the Si--O correlation 
function. 
}
\label{F6}
\end{figure}

Figure \ref{F6} shows the partial structure factor {\FSO} 
obtained by numerically integrating 
Eq.~(\ref{E9}) for the first three individual shells 
(for $\alpha$=1 to 3) using the $R_{ij}^\alpha$ values 
listed in Table \ref{tab3}. 
The results show that shell 2 produces maxima at 
1.8 and 3.52~{\AAI}, and shell 3 produces maxima 
at 1.2, 2.3, and 3.32 {\AAI}. These values are 
quite close to the estimated values of the maxima 
listed in Table~\ref{tab2} from the respective 
shells. Similarly, the minima from shell 2 and 3 in 
Fig.~\ref{F6} correspond to 2.68 {\AAI}, and 1.76, 2.8 and 
3.8 {\AAI}, respectively. 
Once again, these values 
are close to the estimated values of 2.58~{\AAI}, and 
1.72, 2.7 and 3.68~{\AAI}, respectively, as listed 
in Table \ref{tab2}.
By contrast, the contribution from shell 1 provides 
a monotonic decrease of the intensity with an 
increasing value of $Q$ near the FSDP. The first 
shell does not produce any maximum in the region 
between 1 to 4 {\AAI} but a minimum at 2.74~{\AAI}, 
which is estimated to be 2.86 {\AAI} in Table~\ref{tab2}. 
The latter is very close to the exact numerical value 
of 2.74~{\AAI}, when one takes into account the 
perturbative correction of about 0.128~{\AAI} for 
this peak~\cite{corr}.

\begin{figure}[ht!]
\includegraphics[width=0.7\columnwidth]{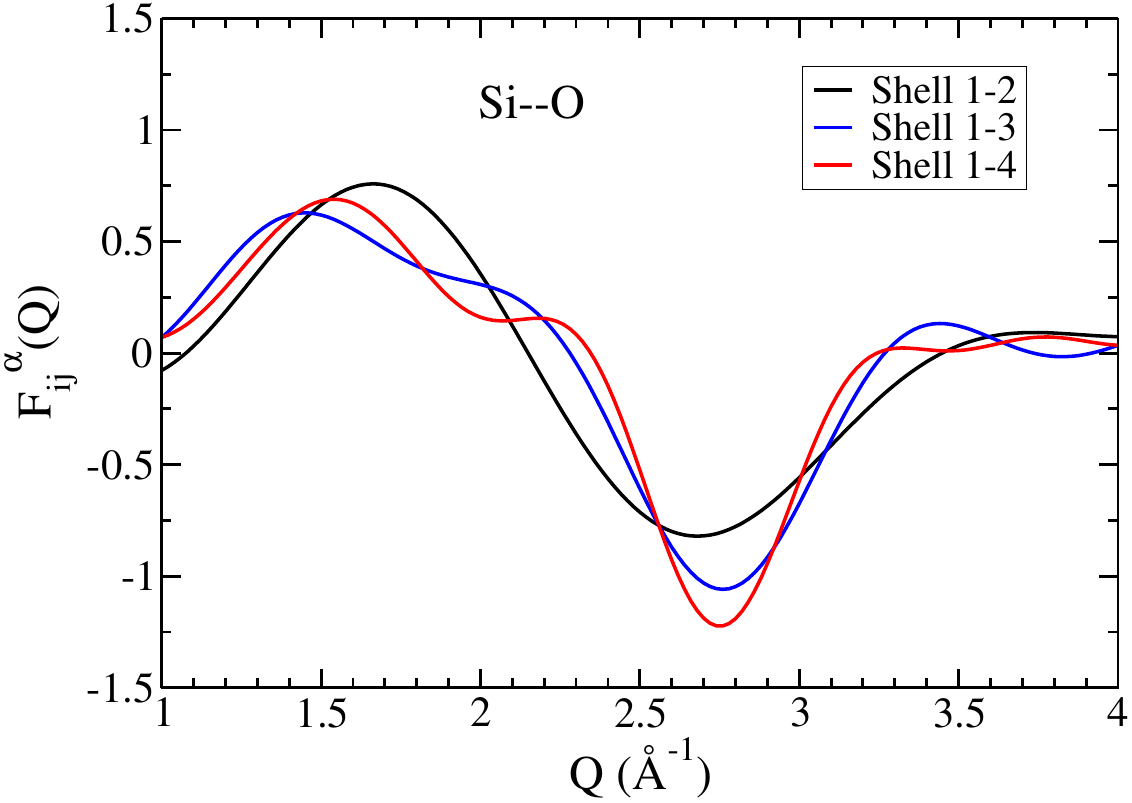}
\caption{
The formation of the FSDP near 1.5 {\AAI} and a 
minimum near 2.8~{\AAI} in {\FSO} originating 
from the first two, first three and first four 
radial shells. The minimum plays an important 
role in the formation of the principal peak 
in Fig.~\ref{F5}. 
}
\label{F7}
\end{figure}  
The observations above suggest that the approximate 
positions of the maxima (peaks) and minima (dips) 
in Fig.~\ref{F6} can be obtained quite accurately 
from the heuristic argument presented in Sec.~IIIC 
using Eq.~(\ref{E8}). These maxima and minima 
collectively lead to the formation of the FSDP 
and the principal minimum in {\FSO}.  
The net effect of the contributions from different 
radial shells to the FSDP and the principal minimum 
is shown in Fig.~\ref{F7}, where {\FSO} obtained 
from the first two, first three and first four 
radial shells are presented. The buildup of the 
FSDP near 1.5 {\AAI} is evident from the plots. 
The resultant {\FSO} obtained from all radial 
shells up to 10.7 {\AA} evolves to produce the 
FSDP near 1.5~{\AAI} and a minimum near 2.8~{\AAI}. 
As stated earlier, there is no principal {\em peak 
or maximum} in {\FSO} in Fig.~\ref{F7} but 
a minimum near 2.8~{\AAI}. This minimum contributes 
negatively to form the principal peak in total $F(Q)$ 
along with the positive contributions from {\FOO} 
and {\FSS} near 2.8~{\AAI}.

\begin{figure}[t!]
\includegraphics[width=0.7\columnwidth]{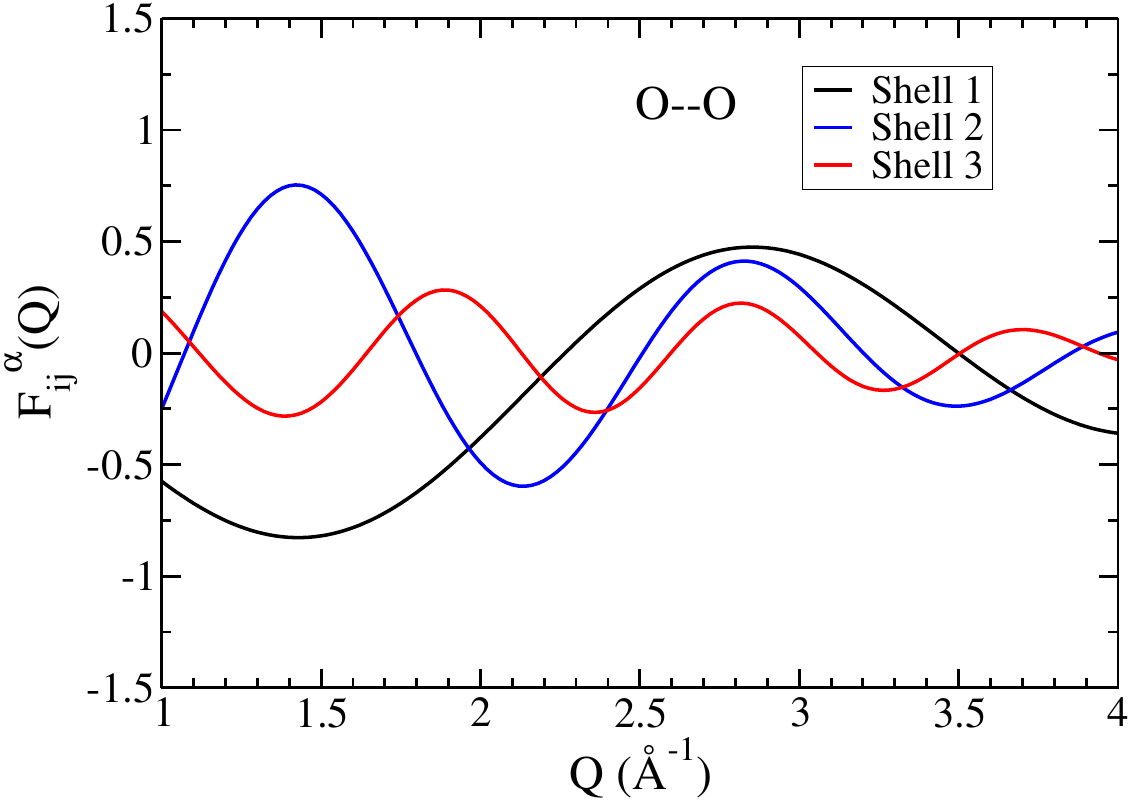}
\caption{
The shell-by-shell contributions from the first three 
radial shells of {\GOO} to {\FOO} for $\alpha$ = 1, 2, 3. 
The maxima associated with the principal peak appear 
near 2.8 {\AAI}. 
}
\label{F8}
\end{figure} 

\begin{figure}[t!]
\includegraphics[width=0.7\columnwidth]{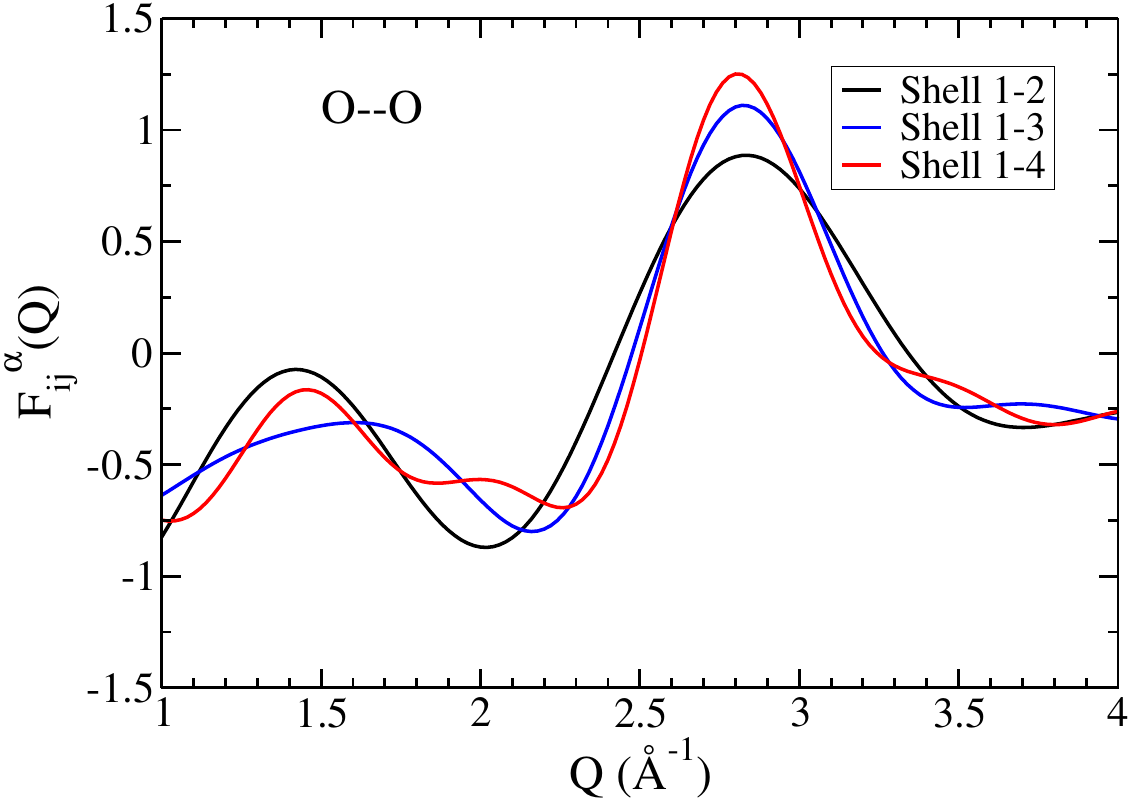}
\caption{
The buildup of the FSDP and the principal peak near 
1.5 and 2.8~{\AAI}, respectively, from small corrections 
originating from the third and fourth radial shells of 
O--O correlations. 
}
\label{F9}
\end{figure}

A similar analysis of O--O correlations reveals that the approximate 
locations of the FSDP and the principal peak in {\FOO} are correctly 
obtained in Sec.~IIIC. Figure \ref{F8} presents the 
contribution from the first three individual radial 
shells of {\GOO}. The results show the presence of 
two maxima at 1.42 and 2.82~{\AAI} originating from 
the second shell. The corresponding estimated 
values of the maxima are found to be 1.51 and 2.72 {\AAI}, 
respectively. Likewise, the first shell and the third shell each 
produces a peak at 2.86 and 2.82 {\AAI}, respectively, which are 
very close to the estimated values of 2.96 and 2.83 {\AAI} 
listed in Table \ref{tab1}. 
%
%
%By contrast, the estimated position of the minimum from the 
%first shell at 1.74 {\AAI} (for $n=1$) considerably deviates 
%from the actual value of about 1.5~{\AAI} observed in Fig.~\ref{F8}. 
%This is not surprising as the numerical values of the maxima 
%and minima are somewhat dependent on the choice of the radial 
%peak value and that the estimated values of the maxima and 
%minima from Eq.~(\ref{E8}) do not include the effect of the 
%finite width of the radial shells by our assumption of the 
%$\delta$-function approximation. 
%
%
The overall effect of the contribution from distant radial 
shells can be seen in Fig.~\ref{F9}, where the intensities 
obtained from the first two, first three and first four 
radial shells of O--O pair correlations are presented. It 
is apparent from Fig.~\ref{F9} that the first two shells 
can produce the intensity and shape of the FSDP and the 
principal peak [in {\FOO}] almost correctly in the 
vicinity of 1.5 and 2.8~{\AAI}, respectively. The contributions from 
the third and fourth shells then refine the shape of 
the intensity curve and add small corrections to the intensity 
and position of the FSDP and the principal peak. 

\begin{figure}[t!]
\includegraphics[width=0.7\columnwidth]{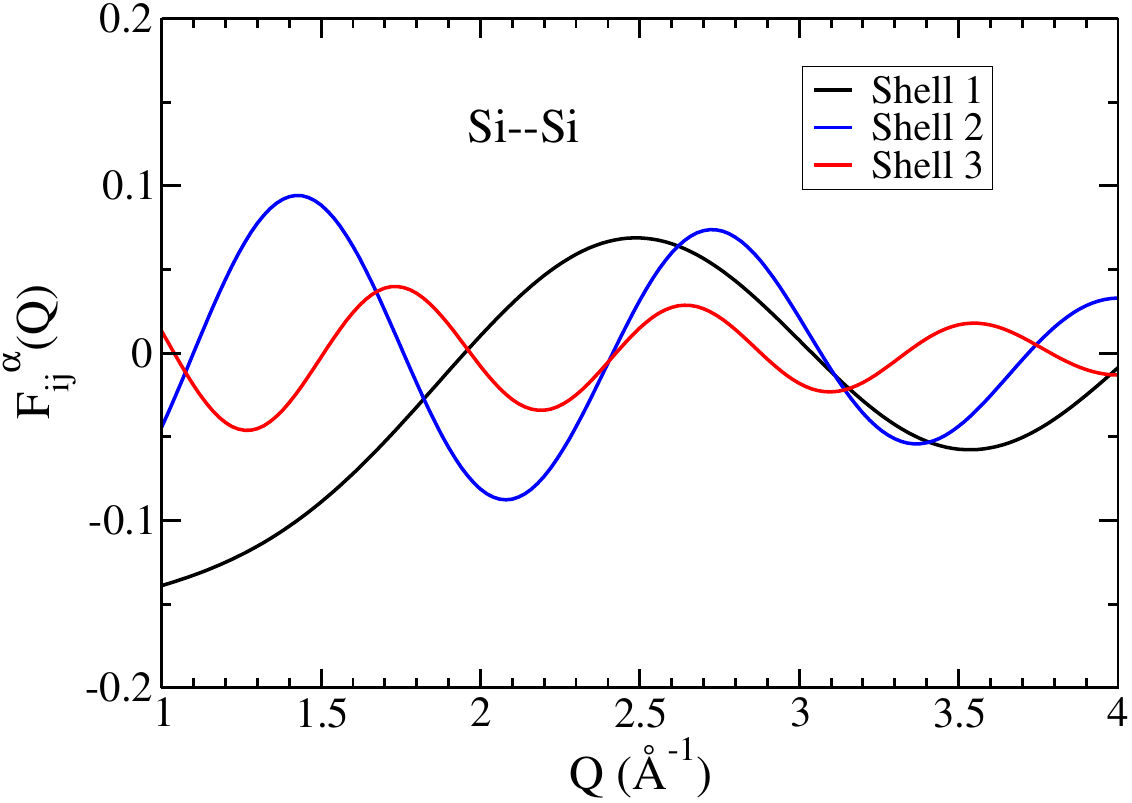}
\caption{
The contribution to the FSDP and the principal peak 
from the first three individual radial shells of 
Si--Si correlations. The radial correlations from 
the second and third shells almost cancel each 
other, leaving behind a small net contribution from 
the first three shells.  
}
\label{F10}
\end{figure}

\begin{figure}[t!]
\includegraphics[width=0.7\columnwidth]{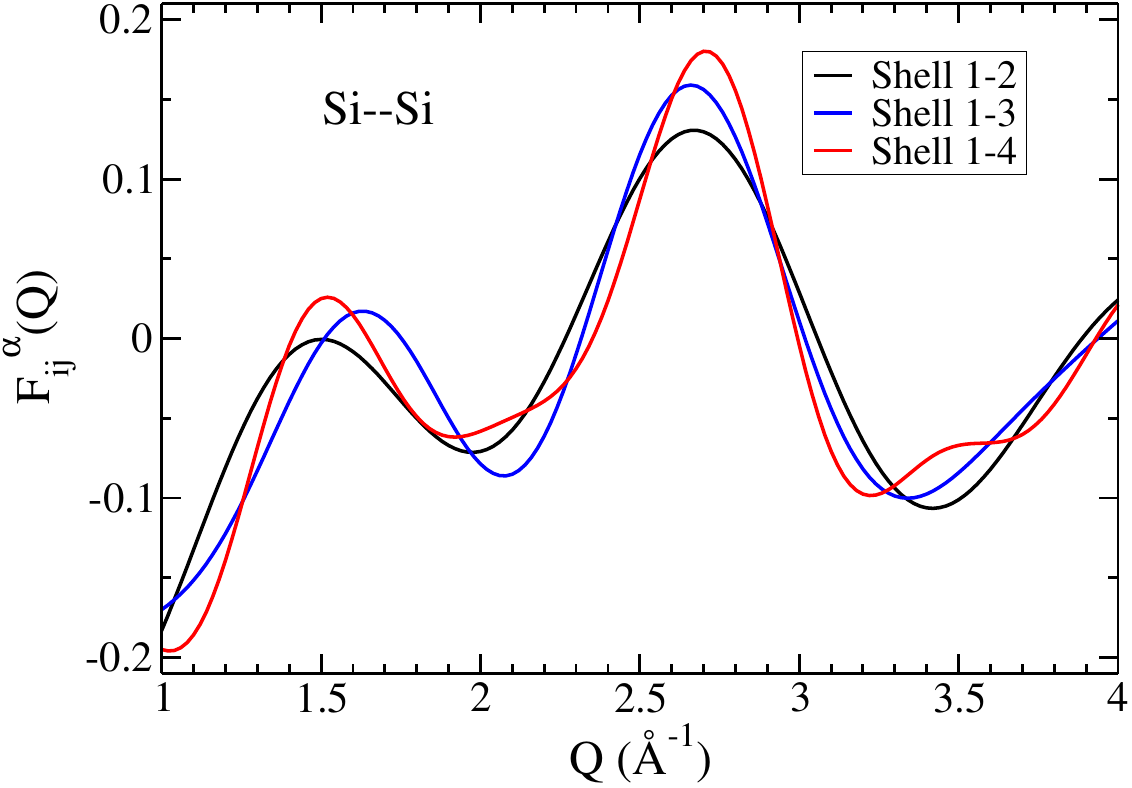}
\caption{
The evolution of the intensity of {\FSS} in the vicinity of 
the FSDP and the principal peak with an increasing number 
of radial shells.  The plots show the contributions from 
the first two, first three, and first four radial shells 
of Si--Si correlations. 
}
\label{F11}
\end{figure}

Similar observations can be made for {\FSS} as well. 
The results for this case are plotted in Figs.~\ref{F10}--\ref{F11}. 
An inspection of the plots in Fig.~\ref{F10} suggests 
that, although the individual contributions from the 
first three shells of {\GSS} are not negligible in 
magnitude, the resultant contribution is quite small near the 
FSDP. This is evident from Fig.~\ref{F11}, where {\FSS} 
is found to be considerably smaller (by a factor of about 10) 
than its Si--O and O--O counterparts in the vicinity 
of the FSDP. The neutron-weighted Si--Si correlations 
thus provide only a small correction to the resultant 
position and the intensity of the FSDP in {\asio2}. 
This statement also applies to the principal peak 
to a lesser extent, where the peak height near 2.7~{\AAI} 
(in Fig.~\ref{F11}) can be seen to increase by a 
small amount as the radial correlations from the third 
and fourth shells are taken into account. 

We conclude this section by making the following remarks.
Firstly, the FSDP in {\FOO} is primarily determined by 
neutron-weighted atomic correlations 
originating from the first two radial shells of {\GOO} 
involving a length scale of about 6 {\AA}. 
The information from the third and fourth shells 
then adds perturbative corrections to the intensity and shape of the FSDP. 
By contrast, the principal peak in {\FOO} involves 
contributions from the first three shells with atomic 
correlations extending up to 9 {\AA}. The inclusion of 
radial information from the fourth shell makes a small 
correction to the height/intensity of the principal peak. 
Secondly, the Si--O correlations from all the four 
shells (extending up to 10 {\AA}) contribute to the 
intensity of the FSDP in {\FSO}. With the exception 
of the first shell, each of the remaining three 
shells produces a peak in the scattering region 
of 1.2--1.85 {\AAI}. These peaks combine with the 
minima from the second, third, and fourth (not 
shown in Fig.~\ref{F6}) shells to form the FSDP 
near 1.5 {\AAI}. The involvement of the first shell is to 
provide a background correction to the FSDP. 
Thirdly, the principal {\em minimum} in {\FSO} in 
the vicinity of 2.8~{\AAI} largely depends on the 
radial correlations from the first three shells, 
which are supplemented by a small correction from 
the fourth shell. 
This minimum effectively cancels out the scattering 
intensity that originates from the neutron-weighted 
O--O pair correlations. The Si--Si correlations then 
provide a small but crucial contribution (of about 2 
part in 10) in the vicinity of the resultant principal 
peak at 2.8~{\AAI}, but its contribution toward the 
intensity of the FSDP is practically negligible 
(cf.~Figs.~\ref{F5} and \ref{F11}).

\subsection{Intensity of the FSDP and the principal peak: A Gaussian approximation}

Having numerically verified the approximate positions of the 
FSDP and the principal peak, and noting the importance of 
contributions from different partial components of the 
structure factor to the FSDP, we now obtain 
an analytical expression for the intensity of $F^{\alpha}_{ij}(Q)$ 
that arises from a 
given shell. Earlier, in Sec.~IIIC, we have shown that the 
approximate position of the peaks in $F^{\alpha}_{ij}(Q)$, 
originating from a given radial shell, can be obtained by placing a 
sufficiently narrow distribution or a $\delta$ function at 
the center of the radial shell. However, the calculation 
of the intensity of the FSDP (or any diffraction peaks in 
general) requires a more sophisticated approach.  To this 
end, the $\delta$-function approximation can be significantly improved 
by replacing the $\delta$ function with a more realistic 
distribution, which can be represented by a linear 
combination of suitable radial basis functions. Assuming that 
the individual radial shells of $g_{ij}(r)$ are Gaussian 
representable~\cite{Dahal:2022}, one can approximate the 
intensity due to the $\alpha$-th shell from Eq.~(\ref{E9}) as
\bea
F^{\alpha}_{ij}(Q) & = & \frac{B_{ij}}{Q}
\int_{R_\alpha} [r\,g_{ij}^{\alpha}(r) - r]\, 
\sin Qr \, dr  \notag \\
& \approx &
\frac{B_{ij}}{Q} \left[\int_{R_\alpha} 
\sum_n rf^{\alpha}_{n}\, \sin Qr \, dr - \int_{R_\alpha}
r \sin Qr\, dr \right] \notag \\
& \approx &
\frac{B_{ij}}{Q} \left[\sum_n I^{\alpha}_{n} - I_0\right], 
\label{E10}
\eea
where $B_{ij} = 4\pi\rho_0\omega_{ij}$, 
the symbol $\int_{R_\alpha}$ indicates that 
the integration is to be carried out from $R_{\alpha}$ to $R_{\alpha + 1}$, 
and $g_{ij}^{\alpha}(r)$ can be expressed as a linear combination 
of Gaussian functions $f^{\alpha}_n(r)$
\[
g_{ij}^{\alpha}(r) = \sum_{n=1}^m f^{\alpha}_{n}(r) 
= \sum_{n=1}^m a_{\alpha n} \, 
e^{-{\displaystyle b_{\alpha n}\,(r - c_{\alpha n})^2}}. 
\]
In general, the radial distribution of atoms in 
the first shell of $g_{ij}(r)$, which is not 
symmetric~\cite{Johnson:1983}, can be well 
approximated by using a few Gaussian functions 
but the distant shells require several Gaussian 
functions in order to produce the distribution accurately within 
the region of $R_{\alpha}$ and $R_{\alpha +1}$ 
with correct boundary conditions.
An analytical expression for $F_{ij}^{\alpha}(Q)$ can 
be obtained by evaluating the integrals in Eq.~(\ref{E10}). 
The first integral is of the type 
\be
I_n = a_n \int_{x_1}^{x_2} \, x \, e^{-b_n(x - c_n)^2}
\, \sin Qx \, dx,  
\label{E11}
\ee
which cannot be evaluated analytically in a 
closed form without the use of error functions.  
However, it is possible to choose a suitable set of 
narrow Gaussian functions, $f^{\alpha}_n$, which 
are characterized by large $b_n$, such that the 
functions decay to zero sufficiently fast before 
they reach the boundary points $x_1$ and 
$x_2$.  This can be achieved by ensuring 
that all $b_n$ satisfy the condition~\cite{mnote} 
\be 
\min[x_2-c_n, c_n-x_1] \gg \frac{1}{\sqrt{b_n}}.  
\label{E12}
\ee
Under this condition, it is possible to obtain an 
analytical expression for $I_n$ in Eq.~(\ref{E11}). 
The calculation of $I_n$ is presented in the 
Appendix. The second integral $I_0$, involving 
$r\sin Qr$, can be analytically calculated. 
The two integrals can be written as 
\be
I_n = a_n \sqrt{\frac{\pi}{4b_n}}\, e^{-\frac{Q^2}{4b_n}} \left[2c_n\sin Qc_n + 
\frac{Q\cos Qc_n}{b_n}\right] 
\label{E13}
\ee 
and 
\be
I_0 = 
\left[\frac{\sin Qr - Qr \cos Qr}{Q^2}\right]_{R_{\alpha}}^{R_{\alpha + 1}}. 
\label{E14}
\ee 
An analytical expression for $F_{ij}^{\alpha}(Q)$ is thus 
obtained by substituting the results from Eqs.~(\ref{E13}) 
and (\ref{E14}) to Eq.~(\ref{E10}) for the $\alpha$-th 
radial shell that ranges from $R_{\alpha}$ to $R_{\alpha+1}$. 
The Gaussian parameters $(a_n, b_n, c_n)$, where $b_n$ satisfies 
the condition in Eq.~(\ref{E12}), for a given $\alpha$ can be 
obtained by fitting the functions with the corresponding 
$g^{\alpha}_{ij}(r)$. It may be noted that 
the Gaussian representation of distant radial 
shells beyond the first shell is highly nontrivial 
due to the presence of sharp cutoffs of $g^{\alpha}_{ij}(r)$ 
with finite values at $r=R_{\alpha}$ and $R_{\alpha+1}$. The 
existence of finite values of $g^{\alpha}_{ij}(r)$ 
at the boundaries require that the Gaussian coefficients 
$b_n$ -- especially those associated with the Gaussians functions 
centered near the boundary -- must be sufficiently large for Eq.~(\ref{E13}) 
to be valid.  In this study, we have employed a constrained 
Monte Carlo approach to obtain the desired values of $b_n$ 
that simultaneously reproduce $g_{ij}^{\alpha}(r)$ 
within the radial region of interest and satisfy the 
inequality in Eq.~({\ref{E12}) for all $n$. 

\begin{figure}[t!]
\includegraphics[width=0.7\columnwidth]{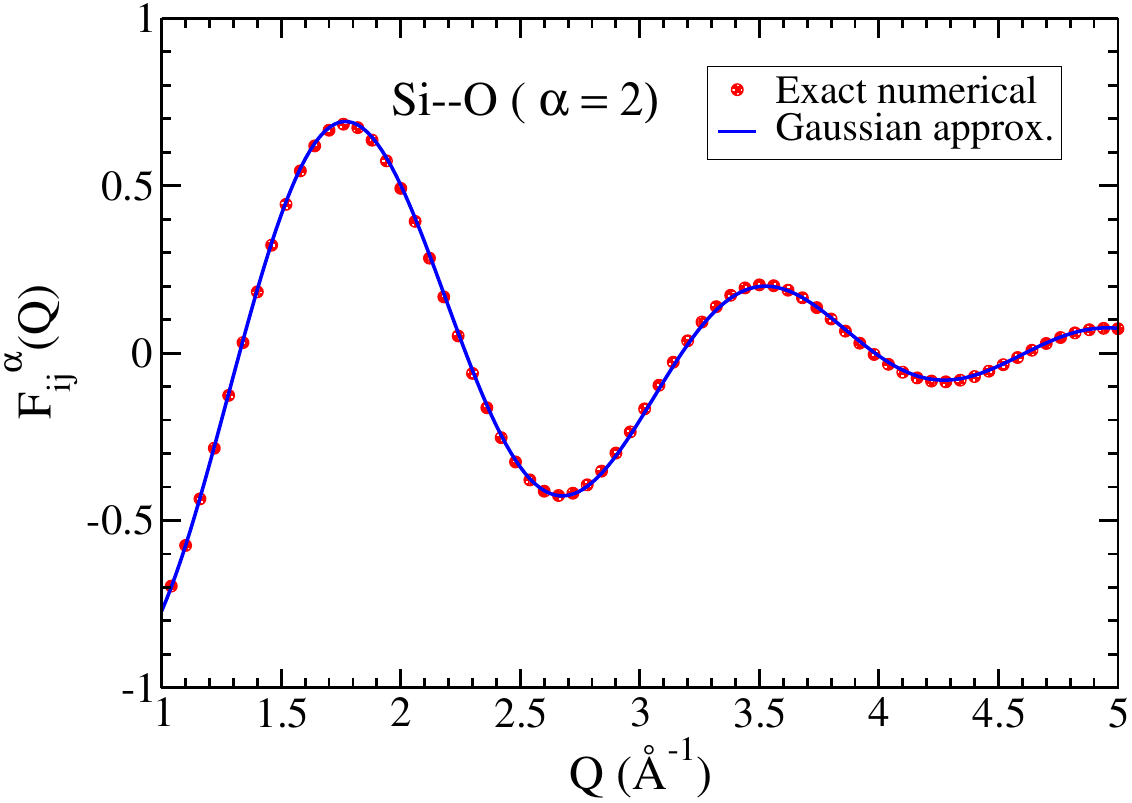}
\caption{
The partial structure factor, $F^{\alpha}_{\text{\scriptsize SiO}}(Q)$, 
for the second shell ($\alpha$ = 2) obtained from exact numerical 
calculations (red circles) and the Gaussian approximation (blue line) 
using Eq.~(\ref{E10}). 
}
\label{F12}
\end{figure}  

\begin{figure}[t!]
\includegraphics[width=0.7\columnwidth]{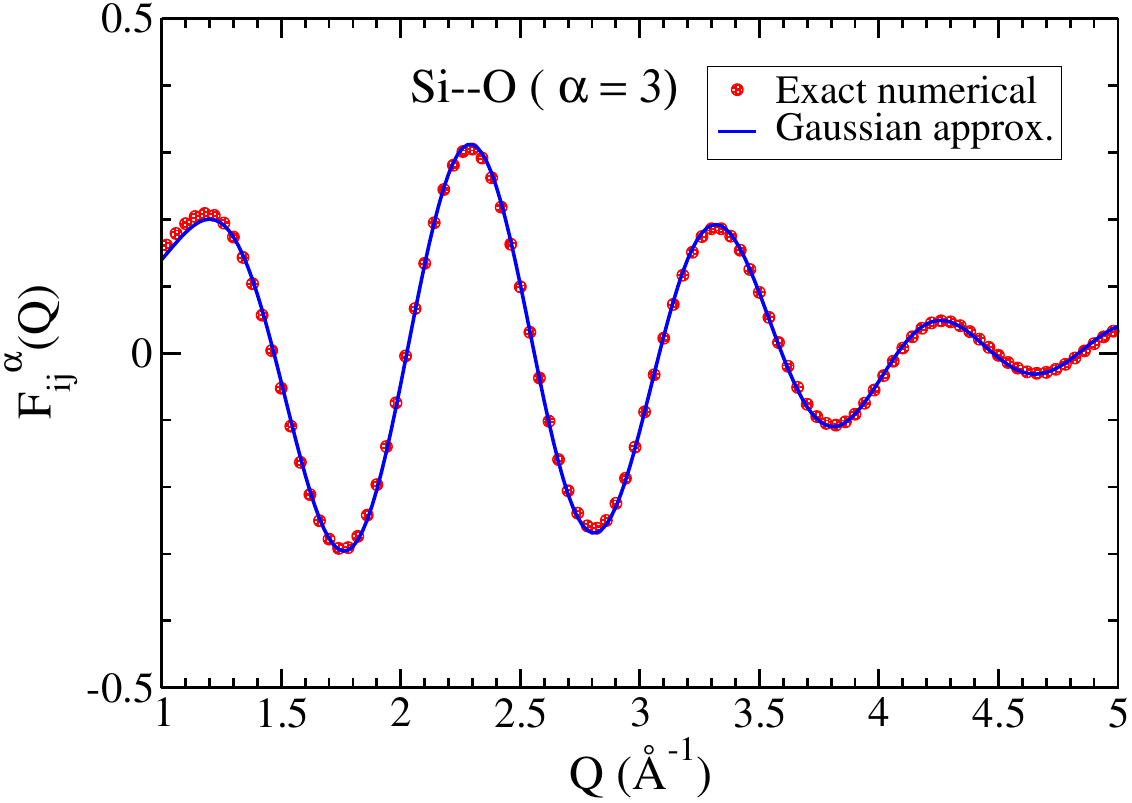}
\caption{
$F^{\alpha}_{\text{\scriptsize SiO}}(Q)$, for $\alpha$ = 3, 
obtained from exact numerical calculations (red circles) 
and the Gaussian approximation (blue line).
}
\label{F13}
\end{figure}  

The efficacy of the Gaussian approximation can be examined 
by comparing the results obtained from Eq.~(\ref{E10}) with 
those from direct numerical calculations using the radial 
range of the shells given in Table \ref{tab3}.  Figures~\ref{F12} 
and \ref{F13} show the results for {\FSO} for 
$\alpha=2$ and 3, respectively. It is evident from the 
figures that the Gaussian approximation can accurately 
reproduce the exact numerical results. The contributions 
to the FSDP and the principal peak are correctly obtained 
from Eq.~(\ref{E10}) for the second and third shells. 
Likewise, the results for the Gaussian approximation of 
{\FOO} in Figs.~\ref{F14} and \ref{F15}, for $\alpha=2$ and 3, 
respectively, closely match with those from direct 
numerical calculations. A similar observation applies to 
Fig.~\ref{F16}, where the intensity plots of {\FSS} for 
$\alpha = 2, 3$ are presented. 
The diffraction intensity obtained from Eq.~(\ref{E10}) 
using the Gaussian approximation of the (partial) 
pair-correlation functions and the positions of the 
FSDP and the principal peak estimated from Eq.~(\ref{E8}) 
provide a complete picture of the origin and structure 
of the FSDP and the principal peak in terms of the 
real-space two-body distributions of silicon and oxygen 
atoms in the radial shells of {\asio2}. 

\begin{figure}[t!]
\includegraphics[width=0.7\columnwidth]{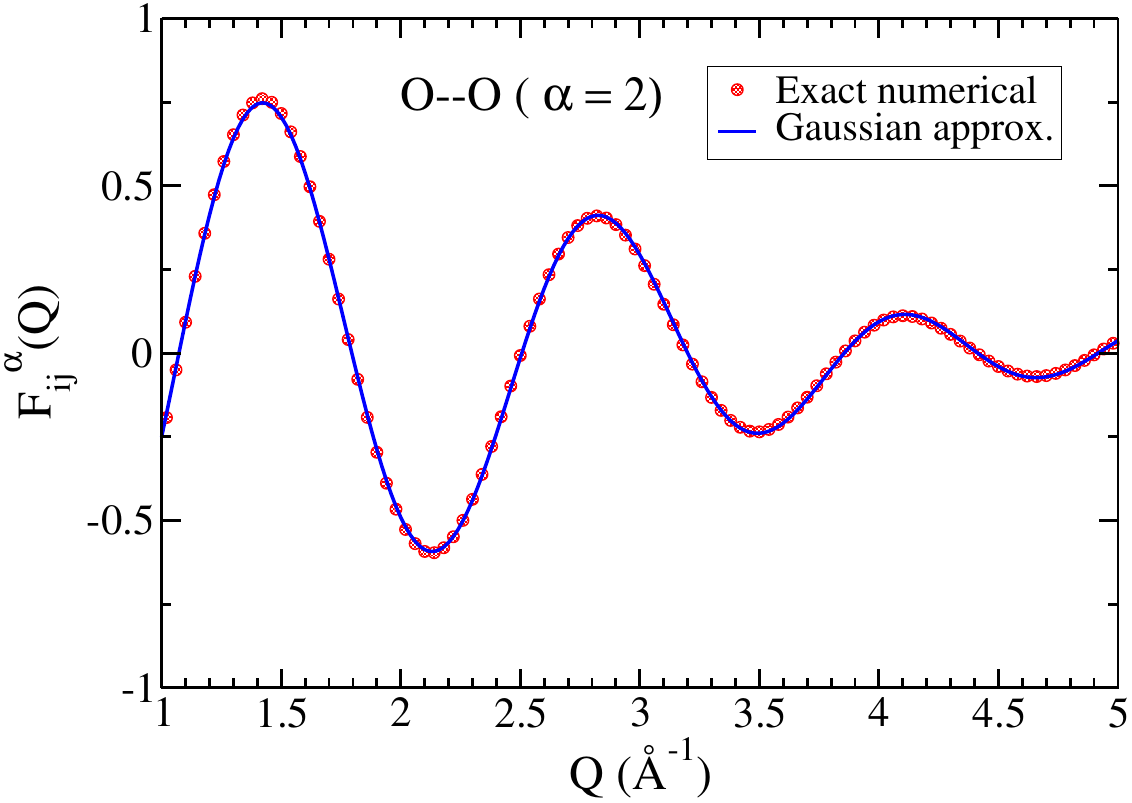}
\caption{
A comparison of $F^{\alpha}_{\text{\scriptsize OO}}(Q)$, 
obtained from exact numerical calculations (red circles), 
with that from the Gaussian approximation (blue line) 
for $\alpha = 2$. 
}
\label{F14}
\end{figure} 

\begin{figure}[t!]
\includegraphics[width=0.6\columnwidth]{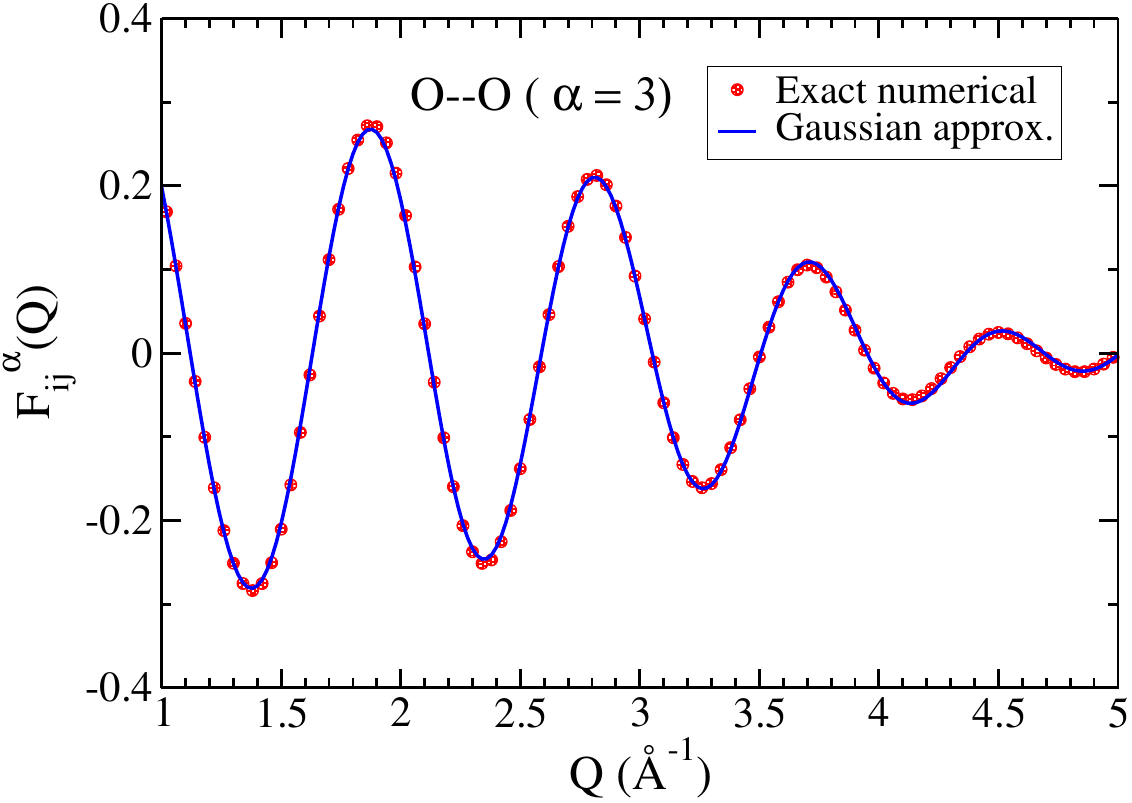}
\caption{
A comparison of the partial structure factor of O--O 
pairs obtained from exact numerical calculations 
(red circles) and that from the Gaussian 
approximation (blue line) for the third shell 
($\alpha = 3$).
}
\label{F15}
\end{figure} 

\begin{figure}[t!]
\includegraphics[width=0.6\columnwidth]{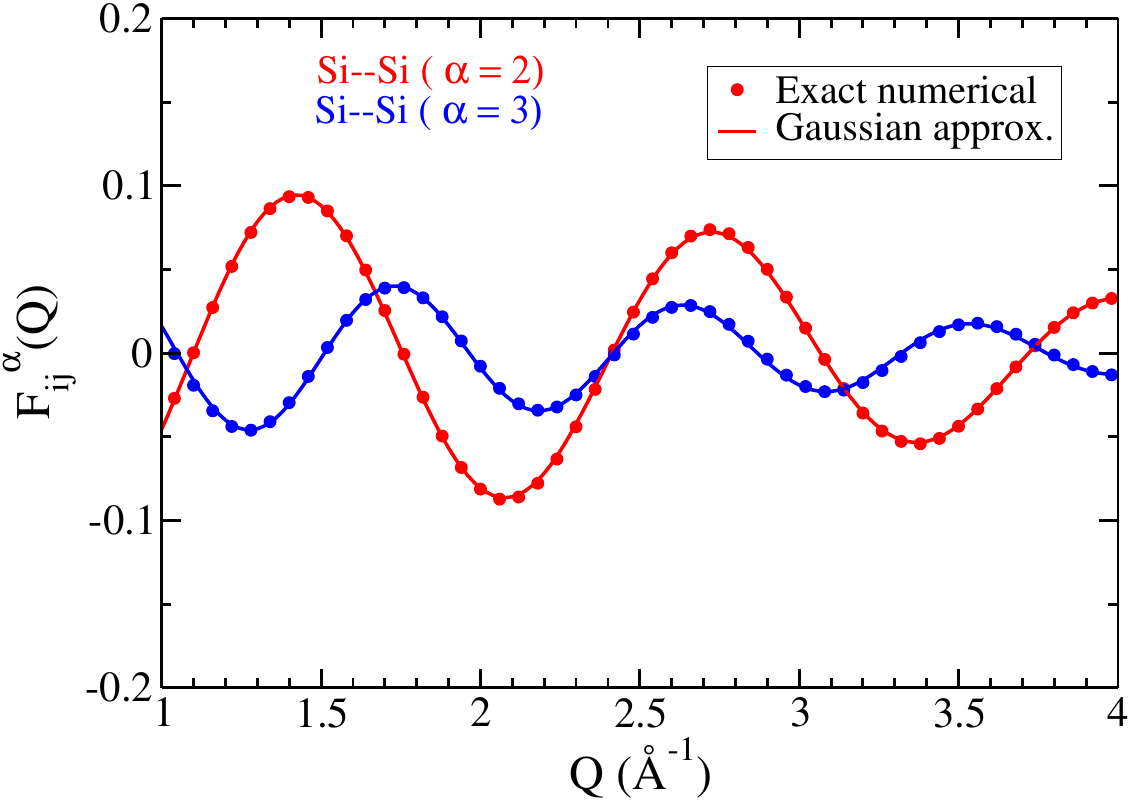}
\caption{
The partial structure factor {\FSS} for the second (red) 
and third (blue) shells obtained from the 
semi-analytical Gaussian approximation (line) and 
exact numerical calculations (filled circle).  
}
\label{F16}
\end{figure}

\section{Conclusions}
In this paper, we have studied the origin and structure 
of the FSDP in {\asio2} with particular emphasis on the 
role of chemical and radial ordering between 
Si--Si, Si--O and O-O pairs. The study leads to the 
following observations. 

An examination of the partial structure factors of 
{\asio2} shows that neutron-weighted Si--O and 
O--O pair correlations on the length scale of 5-8 {\AA} contribute mostly 
to the formation of the FSDP, whereas Si--Si correlations 
(from the third and fourth shells) provide only small 
corrections to the peak position and the intensity in 
the vicinity of the peak. The latter is consistent with 
the relatively weak neutron-weighted Si--Si pair 
correlations, which are found to be approximately ten 
times smaller than the corresponding values of Si--O 
and O--O correlations in the respective PCFs. 

An approximate relationship between the position of the 
FSDP (in the partial structure factors) and the radial 
peak positions (in the corresponding partial PCFs) is obtained 
from the saddle-point approximation of the structure-factor 
integral involving $G_{ij}(r)$. The results from this 
approximation are verified using numerical and semi-analytical 
calculations. A natural corollary of the 
results is that the FSDP is found to be sensitive to the 
pair correlations arising from a few 
specific radial shells. In particular, the first two 
shells of O--O pairs, extending as far as 6 {\AA}, and 
the first three shells of Si--O pairs, with atomic 
correlations of up to 7 {\AA}, are observed to play a 
significant role in forming the position and the 
intensity of the FSDP in {\asio2}. 
By contrast, the shape 
of the intensity curve near the FSDP 
is perturbatively modified by radial correlations 
originating from the remaining shells of Si--O and O--O pairs. 
These results are consistent with those obtained from 
using the continuous wavelet transform~\cite{Uchino:2005} 
of the PCFs showing that Si--O--Si--O (i.e, the second 
shell of Si--O pairs) and O--Si--O--Si--O (i.e., the 
second shell of O--O pairs) correlations 
contribute significantly in the formation of the FSDP 
in {\asio2}. 
Likewise, the principal peak is found to depend mostly 
on the contribution from the Si--O and O--O pairs, 
and a small contribution from Si--Si pairs.  The 
positive contribution from the O--O pairs and 
the negative contribution from the Si--O pairs 
near the principal peak competes with each other and 
largely cancels out. The final position and intensity 
of the principal peak is thus determined by the 
resultant contribution of Si--O and O--O pairs, 
and a small but nontrivial contribution of Si--Si pairs.  

The study yields an accurate semi-analytical 
expression for the diffraction intensity of 
a binary glass originating from a given radial 
shell of the partial PCFs using a Gaussian 
approximation of the latter. The validity of 
the approximation is confirmed by directly comparing 
the results with those from exact numerical calculations. 
The approximation is independent of the nature of 
amorphous solids and it can be applied to both 
elemental and multi-component systems. 
The expression is found to be particularly 
useful for a complete characterization of 
the contributions emanating from each radial 
shell (of the PCFs) to the FSDP, and to 
obtain quantitative values of relevant 
radial length scales that form the basis 
of medium-range ordering in {\asio2}. 
The results can be readily generalized and 
employed to other network glasses. 

\appendix*
\section{Approximation of $\int_{a}^{b} rg(r)\, \sin Qr\, dr$}
In this section, we obtain an approximate analytical 
expression for the integral above when the function 
$g(r)$ is Gaussian representable and it satisfies a 
certain condition. Expressing $g(r)$ as a linear 
combination of Gaussian functions, the integral can 
be written as 
\be 
I = \sum_{i=1}^n \int_a^{b} r a_i \, \, e^{-b_i(r - c_i)^2} 
\sin Qr \; dr = \sum_{i=1}^{n} a_i I_i, \notag 
\label{A1}
\ee
where the coefficients $(a_i, b_i, c_i)$ parameterize the 
Gaussian basis sets. Substituting $y=r-c_i$ and 
expanding $\sin[Q(y+c_i)]$, one obtains 
\[
I_i = I_i^{(1)} +  I_i^{(2)} + I_i^{(3)} + I_i^{(4)}, 
\]
where 

\bea
I_i^{(1)} &=& \sin Qc_i \int_{r_1}^{r_2} y \, e^{-b_iy^2} \,\cos Qy \, dy, 
\notag \\ 
I_i^{(2)} &=& \cos Qc_i \int_{r_1}^{r_2} y \, e^{-b_iy^2} \sin Qy \, dy, \notag \\
I_i^{(3)} &=& c_i \sin Qc_i \int_{r_1}^{r_2} e^{-b_iy^2} \cos Qy \, dy, \notag \\
I_i^{(4)} &=& c_i \cos Qc_i \int_{r_1}^{r_2} e^{-b_iy^2} \sin Qy \, dy. 
\label{A2}
\eea 
The upper and lower limits of the integrals in (\ref{A2}) 
are given by $r_1=a-c_i$ and $r_2=b-c_i$, respectively.  In 
general, the integrals in Eq.~(\ref{A2}) cannot 
be expressed in a closed form without using error functions. 
However, if the value of $b_i$ can be chosen in such a way 
that $b_i r_0^2 \gg 1$, where $r_0=\min(|a-c_i|, |b-c_i|)$, 
then the function $\exp(-b_i y^2)$ inside the 
integrals decays sufficiently rapidly within the 
interval $[a-c_i, b-c_i]$ and the upper and lower limits 
of the integrals can be replaced by +$\infty$ 
and -$\infty$, respectively. Under this condition, the 
odd integrals vanish and one is left with 
\bea
I_i & = & I_i^{(2)} + I_i^{(3)} \approx \cos Qc_i \int_{-\infty}^{\infty} y 
\, e^{-b_iy^2} \sin Qy \, dy \notag \\
& & +  \, c_i \sin Qc_i \int_{-\infty}^{\infty} e^{-b_iy^2} \cos Qy \, dy. 
\label{A3}
\eea

Equation \ref{A3} involves two standard integrals, which 
are given by 
\[ 
\int_{-\infty}^{\infty} e^{-b_iy^2} \cos Qy \, dy = 
\sqrt{\frac{\pi}{b_i}}\, e^{-\frac{Q^2}{4b_i}}
\] 
and 
\[ 
\int_{-\infty}^{\infty} y e^{-b_iy^2} \sin Qy \, dy = 
\frac{Q}{2b_i}\,\sqrt{\frac{\pi}{b_i}}\, e^{-\frac{Q^2}{4b_i}}.  
\]
Collecting all the results, the original integral $I$ can 
be approximated as 
\be
I \approx \sum_i^n a_i\, \sqrt{\frac{\pi}{4b_i}}\, e^{-\frac{Q^2}{4b_i}} \left[2 c_i \sin Qc_i + \frac{Q\cos Qc_i}{b_i} \right]. 
\label{A4}
\ee
Writing $b_i = 1/2\sigma_i^2$, the condition of validity of the result 
in Eq.~(\ref{A4}) can be stated as 
$\sqrt{2} \sigma_i \ll r_0 = \min(|a-c_i|,|b-c_i|)$ for all 
$b_i$, $c_i$, and $(a,b)$. The validity of this 
expression has been verified by numerical experiments. 

\bibliography{silica.bib}

\begin{thebibliography}{57}
\bibitem [{pea()}]{peak}%
  \BibitemOpen
  \href@noop {} {}\bibinfo {note} {Assuming $g(r)$ is Gaussian representable,
  the peak positions in $g(r)$ and $r^2g(r)$ are separated by $\delta r_n
  \approx 2\sigma_n^2/r_n$, where $r_n$ and $\sigma_n$ are the Gaussian peak
  and width of the $n$-th radial shell of $g(r)$, respectively. Thus, the
  approximation holds for $\sigma_n \ll r_n$, which is always satisfied by a
  well-defined radial shell. The presence of linear and quadratic shifts of $r$
  in $G(r)$ and $rG(r)$, respectively, slightly modify this
  estimate.}\BibitemShut {Stop}%
\bibitem [{cor()}]{corr}%
  \BibitemOpen
  \href@noop {} {}\bibinfo {note} {The maximal value of the perturbative
  correction is obtained for $n$ = 1 and $r_{\text{\tiny Si-O}}$ = 1.65~{\AA}
  in Eq.~(\ref{E7}). This gives a value of 0.077~{\AAI} for a maximum and
  0.128~{\AAI} for a minimum. For distant radial peaks and $n>$ 1, the
  correction can be ignored.}\BibitemShut {Stop}%
\bibitem [{mno()}]{mnote}%
  \BibitemOpen
  \href@noop {} {}\bibinfo {note} {Noting that a Gaussian function decays to
  almost zero at a distance of $4\sigma$ from the center, a sharper upper bound
  for $\sigma^2$ (=$1/2b$) can be stated as $4\sigma \le \min[|x_2-c|,
  |x_1-c|]$.}\BibitemShut {Stop}%
\end{thebibliography}

\end{document}